\documentclass{article}
\usepackage{amsmath,amsfonts,amssymb}

\usepackage{xcolor}
\usepackage{graphicx}
\usepackage{subcaption}

\usepackage[bookmarks,
			colorlinks = true,
            linkcolor = blue,
            urlcolor  = blue,
            citecolor = blue,
            anchorcolor = blue]{hyperref}
\usepackage{apacite}

\usepackage{lineno}
\usepackage[affil-it]{authblk}

\title{Conditional GLMMs for reaction times in choice tasks}

\author[a,*]{Mauricio Tejo}
\author[b]{Cristian Meza} 
\author[c]{Fernando Marmolejo-Ramos}

\affil[a]{CIMFAV - Instituto de Estadística, Universidad de Valparaíso, Chile.}
\affil[b]{CIMFAV - Instituto de Ingeniería Matemática, Universidad de Valparaíso, Chile.}
\affil[c]{College of Human Sciences and Culture, Flinders University, Australia.}
\affil[*]{Corresponding author: mauricio.tejo@uv.cl}
\begin{document}

\date{}

\maketitle

{
\begin{abstract}

.
This study connects two methods for modeling reaction times (RTs) in choice tasks: (1) the first-hitting time of a simple diffusion model with a single barrier, representing the cognitive process leading to a response, and (2) Generalized Linear Mixed Models (GLMMs). We achieve this by analyzing the RT distributions conditioned on each response alternative. Because certain diffusion model variants yield Inverse Gaussian (IG) and Gamma distributions for first-hitting times, we can justify using these distributions in RT models. Conversely, employing IG and Gamma distributions within GLMMs allows us to infer the underlying cognitive processes, reconstructed under a moment-based approximation, being this reconstruction exact when random-effect conditionals are set at $C=0$. We demonstrate this concept through simulations and apply it to previously published real-world data. Finally, we discuss the scope and potential extensions of our approach.

\end{abstract}
}

\newpage
\section{Introduction}\label{sec:sta}

\subsection{Background}

In experimental psychology, reaction times (RTs)—the time between a stimulus and a behavioral response—have long been a central focus of study, aiming to understand the processes underlying their distribution [e.g., \shortciteA{woodrow1911reaction}, \shortciteA{smith1968choice}, \shortciteA{wainer1977speed}, \shortciteA{ratcliff1978theory}, \shortciteA{meyer1988modern}, \shortciteA{posner2005timing}, \shortciteA{matzke2009psychological}, \shortciteA{baayen2010analyzing}, \shortciteA{balota2011moving}, \shortciteA{woods2015factors}, \shortciteA{donkin2018response}, \shortciteA{de2019overview}, \shortciteA{tejo2019theoretical}, \shortciteA{rousselet2020reaction}]. RT models can be broadly classified (not mutually exclusively) as: (1) quantitative distribution measurements, focused on capturing key features of empirical RT distributions; and/or (2) theoretical models, deriving probability distributions based on cognitive mechanisms [e.g., \shortciteA{matzke2009psychological}, \shortciteA{ratcliff2011diffusion}, \shortciteA{anders2016shifted}, \shortciteA{tejo2019theoretical}].

A common example of a quantitative RT distribution model is the Ex-Gaussian (EG) distribution, which often provides a good fit to empirical RT data. However, its parameters lack a clear cognitive interpretation [e.g., \shortciteA{ratcliff1978theory}, \shortciteA{rohrer1994analysis}, \shortciteA{matzke2009psychological}, \shortciteA{palmer2011shapes}, \shortciteA{marmolejo2013power}, \shortciteA{marmolejo2015automatic}, \shortciteA{marmolejo2015efficacy}, \shortciteA{velez2015new}, \shortciteA{osmon2018non}]. Conversely, theoretical models aim to derive models based on the cognitive mechanisms underlying responses in choice tests, with model complexity depending on the complexity of the experimental task. The interest here is on modeling RTs as probability distributions based on cognitive mechanisms.

Experimental task complexity varies. In simple-choice (SC) tasks, stimuli are presented one at a time, and participants make one of two possible responses to categorize the stimulus (e.g., a participant views a picture and classifies it as either positive or negative in valence). In multiple-choice (MC) tasks, stimuli are also presented one at a time, but participants must select one of three or more possible responses to categorize the stimulus (e.g., a participant views a picture and classifies it as positive, neutral, or negative in valence). (Here, we assume that individuals are compelled to choose and respond with one of the available options provided in the task.) For SC tasks, theoretical models typically conceptualize the underlying cognitive mechanisms experienced by participants during the task. These models are then used to calculate the first instance at which one of the two response options is emitted.

%%%%%%%%%%%%%%%%%%%%%%%%%%%%%%%%%%%%%%%%%%%%%%%%%%%%%%%
\subsection{Preliminaries}

Regarding the underlying cognitive mechanism involved in a SC task, a general scheme of a cognitive process, denoted as  $\{X_{t}\}_{t\in\mathbb{R}_{+}}$, may consider a first-time step $\theta\geq0$, during which the individual primarily receives the stimulus and no response can be produced during this time interval. Following this, the decision stage begins by processing the information related to the task's difficulty along with random fluctuations. A response is finally produced when $X_{t}$  reaches one of the two  barriers, say 0 or $b$ (with $0<X_{\theta}<b$), for the first time. The distribution of this first-hitting time corresponds to the RT distribution. Depending on the assumed dynamics of  $\{X_{t}\}_{t\in\mathbb{R}_{+}}$, calculating this first-hitting time can be complex [see e.g. \shortciteA{laberge1962recruitment}, \shortciteA{ratcliff1978theory}, \shortciteA{ratcliff2002estimating}, \shortciteA{usher2002hick}, \shortciteA{horrocks2004modeling}, \shortciteA{brown2008simplest}, \shortciteA{blurton2017first}].

A simplified single-barrier model remains useful as a representation of the cognitive process leading to a specific response. By calculating the RT distribution conditioned on each response alternative, the overall RT distribution can be obtained as a mixture of these conditional distributions, weighted by the marginal probabilities of each response (via the Law of Total Probability, LTP). This principle can be extended to MC tasks.

A simple mathematical representation of the cognitive process in a choice task can be expressed as:
\begin{equation}\label{eq:dif}
X_t=X_\theta-\nu(t-\theta)+e_{t-\theta},
\end{equation}
for $t\geq\theta\in\mathbb{R}_{+}$, and $X_t=X_\theta$ otherwise. Here, $X_\theta>0$ represents the initial processing state, with $\theta$ also known as the ``non-decision parameter''; $\nu$ is the ``mean rate of information accumulation'' (influenced by individual differences in information processing or stimulus characteristics reflecting task difficulty), and $\{e_{t}\}_{t\in\mathbb{R}_{+}}$ is a stochastic process representing random fluctuations, typically assumed to be a Wiener or Brownian motion process [\shortciteA{ratcliff2011diffusion}, \shortciteA{tejo2019theoretical}].  It is well-known that when $X_\theta$  and $\nu$ are assumed to be fixed parameters, such that $\nu>0$, the resulting distribution of the one-barrier first-hitting time after $\theta$  of $X_t$  to 0, 
\begin{equation}\label{stopping}
\tau_0:=\inf\{t>\theta: X_t=0\},
\end{equation}
is a shifted version of the Inverse Gaussian (IG) or Wald distribution. Other distributions within the Gamma and Birnbaum-Saunders (BS) families can be derived for this first-hitting time distribution by modifying assumptions about the components of Equation \eqref{eq:dif} [\shortciteA{jackson2009randomization}, \shortciteA{tejo2018fatigue}, \shortciteA{tejo2019theoretical}]. These resulting distributions serve as simple theoretical RT models that have demonstrated good quantitative fits to real RT data [\shortciteA{anders2016shifted}, \shortciteA{tejo2018fatigue}, \shortciteA{tejo2019theoretical}]. However, the specific parameter values will depend on experimental factors, such as the stimuli and their corresponding responses. These experiments require a sufficient number of replications per factor (discussed further later).

Similarly, a connection can be established between the parameters of a set of RTs (e.g., mean and variance) and the parameters of the underlying diffusion process (e.g., starting processing point and mean rate of information accumulation). In other words, given a random sample of RTs and estimating its parameters, we can reconstruct a cognitive scheme like \eqref{eq:dif} such that $\tau_0$ (the first hitting time/RT) follows a distribution similar to the original random sample. This is a key focus of this work.

We support the foundation that the validity of a proposed model depends on its performance as a quantitative distribution measurement and its theoretical grounding. Given the varying complexity of choice tasks based on stimuli and responses, a general RT model should incorporate a more complex structure that explicitly considers different stimuli and individual effects as covariates. Generalized Linear Mixed Models (GLMMs) provide such a hierarchical structure [\shortciteA{mcculloch2005generalized}], linking mean RTs with covariates that specify the relationship between RTs and the applied stimuli. In this framework, the dependent variable is RT, with each stimulus (or stimulus combination) treated as a “fixed effect” at various levels, and “random effects” accounting for individual variability across trials [\shortciteA{tuerlinckx2006statistical}, \shortciteA{lo2015transform}, \shortciteA{stroup2012generalized}, \shortciteA{molenaar2015bivariate}, \shortciteA{de2019overview}]. However, in this work, following the LTP, we will use GLMMs to model the RT distribution conditioned on each alternative response.

Often, responses are classified as correct or incorrect based on their congruence with the presented stimuli [e.g., \shortciteA{de2019overview}]; in these cases, the two-barrier model represents a choice between a correct and an incorrect response. However, this classification is not always straightforward. For instance, \shortciteA{marmolejo2020your} presented participants with a spectrum of facial expressions, from happiest to saddest, asking them to classify each as “happy” or “sad.” “Intermediate” expressions were essentially neutral (neither happy nor sad), making a clear “correct”/“incorrect” classification difficult. This is even more problematic with diverse stimuli like landscapes, where similar categorization is required. Therefore, it is generally more useful to analyze the responses themselves, examining their frequency and dispersion at each stimulus level. When responses can be reliably categorized as “correct” or “incorrect,” a more in-depth analysis can determine which stimulus levels lead to higher error rates.

Candidate models for the RT distribution, conditioned on each possible response, can be justified by the theoretical arguments concerning one-barrier first-hitting time distributions. Some resulting distributions, such as the IG and Gamma, are suitable for GLMMs as they belong to the Exponential Family. The BS distribution, closely related to the IG distribution [\shortciteA{desmond1986relationship}, \shortciteA{balakrishnan2009mixture}, \shortciteA{leiva2015modeling}], has been recently proposed as a theoretical model for simple-choice RTs [\shortciteA{ranger2015race}, \shortciteA{tejo2018fatigue}, \shortciteA{tejo2019theoretical}], but is not directly supported within standard GLMMs. While \shortciteA{villegas2011birnbaum} developed a constructive linear mixed model similar to GLMMs for the logarithm of the dependent variable, we will not consider the BS distribution here. This is because transforming the RTs distorts their ratio scale properties [\shortciteA{lo2015transform}], and transforming back to the original RT scale to infer the underlying cognitive process introduces transformations and approximations that could lead to significant errors. Furthermore, deriving BS-distributed stopping times from a diffusion process like \eqref{eq:dif} is an approximation, unlike the exact derivation for IG-distributed stopping times [see \shortciteA{tejo2018fatigue}]. 

\textit{Remark.} Another frequently used model for RTs that can be grounded in cognitive mechanisms is the Ex-Wald distribution, obtained as the convolution of an Exponential with a Wald (or Inverse Gaussian) distribution [e.g., \shortciteA{schwarz2001ex}, \shortciteA{schwarz2002convolution}, \shortciteA{heathcote2004fitting}]. Nevertheless, this model also falls outside our GLMM framework because it is not a member of the Exponential Family: the Ex-Wald’s density involves products of exponential terms, complementary error functions (or normal CDFs), and parameter-dependent integrals generated by the convolution.

To obtain the complete RT distribution, we must also consider the distribution of responses. These can be consistently estimated via frequencies based on stimuli/trials, although a more structured modeling can be developed (we will discuss this later). As previously mentioned, the LTP allows us to obtain the overall RT distribution: if $Y$ represents the RT random variable for a given experiment, and $R_1$, $R_2$,\ldots, $R_L$ are the possible responses, then the cumulative distribution function of $Y$ is given by 
\[
F(y):=P(Y\leq y)=\sum_{l=1}^{L}P(Y\leq y|R_l)P(R_l).
\]
Thus, the distribution of $Y$ is a mixture of conditional distributions, weighted by the probabilities of the corresponding responses. 

Building on the preceding discussion, the next section details our modeling approach and presents simulations demonstrating the reconstruction of \eqref{eq:dif} for IG and Gamma RTs. Section \ref{sec:experiment} applies our methodology to a previously published experiment, and Section \ref{sec:discussion} discusses the scope of our proposal and potential extensions.

%%%%%%%%%%%%%%%%%%%%%%%%%%%%%%%%%%%%%%%%%%%%%%%%%%%%%
\section{Model approach}\label{sec:modelapp}

We begin by defining a hierarchical structure for the dependent variable (RTs) in a choice task where individuals are presented with various types of stimuli at different levels and asked to choose one of several alternatives as quickly as possible. For illustration, consider two response alternatives, $R_1$ and $R_2$, and different types of stimuli (referred to as “fixed effects”) at various levels. Let $n$ be the total number of combined levels across all stimulus types. Let $Y_{ijk}$ denote the $j$-th RT of the $k$-th individual under the $i$-th level, where $j=1,\ldots,n_i$, $k=1,\ldots ,m$ and $i=1,\ldots,n$. 

The objective is to estimate the RT distributions by focusing on the conditional distributions $Y|R_l$, where $l=1,2$ and $Y=\{Y_{ijk}\}_{ijk}$. Notice that it can be extended to multiple alternatives ($l=1,2,\ldots, L$). Conditional random variables operate on a reduced sample space defined by the conditioning event.  In practice, this translates to partitioning the dataset into subsets corresponding to each response outcome. Thus, we empirically split our RT data based on each response. While we model $Y$ conditional on each response, for notational simplicity, we omit the explicit conditional notation. Therefore, we use $Y$ instead of  $Y|R_l$, with the understanding that we are implicitly modeling the conditional distribution.

In this experimental task, each individual's RT result is represented as a random vector of RTs. This vector is denoted as
\[
Y_k=(Y_{11k},\ldots,Y_{1n_1k},\ldots,Y_{n1k},\ldots,Y_{nn_nk})^\top,
\]
and captures the RTs across different stimuli, their levels, and replications. Here, $\top$ indicates the transpose operation. 

We consider a random sample $Y_1,\ldots ,Y_m$, where each $Y_k$ is an independent and identically distributed (IID) random element with a cumulative distribution function $F$. Assuming $F$ has a probability density function (PDF) $f$ belonging to the Exponential Family, we propose a hierarchical GLMM structure for the mean RTs:
\begin{equation}\label{eq:hiestruc}
\mu_{ijk}:=E(Y_{ijk}|C_{k}=c_{k})\text{ and }h(\mu_{ijk})=x_{ij}^{\top}\beta+c_{k},
\end{equation}
where $C_{k}$ represents a random effect acting as a random intercept for each individual; $h$ denotes the ``link function'', $\beta=(\beta_1,\ldots,\beta_n)^\top$ is a fixed-effect vector of unknown regression parameters, and $x_{ij}=(x_{ij1},\ldots,x_{ijn})^\top$ is a vector containing covariate values associated with $\beta$. In this context, we treat stimuli or fixed effects as categorical variables.  The subscript $i$ encompasses all stimuli types and their levels. Consequently, the $x_{ij}$ vectors can be expressed as $x_{ij}=(\delta_{i1},\ldots, \delta_{in})^\top$ for all $j$. Here, $\delta_{ii'}$ is an indicator function: it is 1 if $i=i'$ and 0 otherwise. This structure ensures that the elements of each $x_{ij}$ vector always sum to 1.

For the random effects, which influence the intercept in the linear predictor, we adopt a simplified approach by considering variations only at the individual level. This means each individual has a unique random intercept. While we acknowledge the potential for incorporating other random effects, such as those related to stimuli types, difficulty levels, and within-individual correlations [e.g., using $\{C_{ijk}\}_{ijk}$ as suggested by \shortciteA{baayen2010analyzing}], we opt for parsimony in this initial presentation. (However, as a robustness check, in addition to these individual-level random intercepts we also fit models with a \emph{participant-level random slope} for the stimulus score, i.e.\ a $(1+\text{stimulus}\mid\text{participant})$ structure that lets the stimulus effect vary across individuals; note that this is a participant-level random slope, not a random effect defined at the stimulus level. The two random-effect structures are compared in Table~\ref{tab:modelcmp} through AIC, BIC and the estimated random-effect variances).
Therefore, we assume that the random effects, denoted as $\{C_{k}\}_{k=1,\ldots,m}$ are IID according to a random variable $C$ with a cumulative distribution function $G$. We further assume that $G$ has a continuous PDF $g$. This simplified random-effects structure allows us to focus on the core aspects of our approach. We leave the exploration of more complex random-effect structures for future work, as discussed in Section \ref{sec:discussion}.

GLMMs inherently exhibit a relationship between the mean and variance.  As described in \shortciteA{tsou2011determining}, this relationship can be expressed as: 
\begin{equation}\label{eq:mean-var}
V(Y_{ijk}|C_{k}=c_{k})=\phi \mu_{ijk}^{\lambda},
\end{equation}
where $\lambda$ is a suitable constant and $\phi>0$ is a parameter dispersion (\textbf{note:} the mean-variance relationship described by Tsou -among others- holds for generalized linear models (GLMs); consequently, for our GLMM -essentially a ``GLM + random effects'' - this relationship applies conditionally on the random effects). Different values of $\lambda$ correspond to different distributional families. However, this power relationship can serve as a ``guide'' for proposing candidate families, but it does not provide a unique identification criterion. For example, the Inverse Gaussian (IG) distribution can be associated with both $\lambda=1$ and $\lambda=3$, as can the Poisson distribution; whereas $\lambda=2$ characterizes the Gamma distribution, as well as the Lognormal and Weibull distributions [\shortciteA{tsou2011determining}].
Since our hierarchical structure is developed conditional on each response, we allow the parameters in Equation \eqref{eq:mean-var} to vary depending on the specific response. However, we maintain the same link function $h$ across all responses, assuming that the conditional distributions $Y|R_l$ belong to the same family. 

As discussed in Section \ref{sec:sta}, the conditional distribution of $Y_{ijk}$ given a specific response can be theoretically modeled as the first-hitting time distribution of a diffusion process, as represented in Equation \eqref{eq:dif}. The specific form of the diffusion process, also detailed in Section \ref{sec:sta}, determines the appropriate link function $h$ for the GLMM. For example, if the diffusion process leads to an IG distribution for the RT, common link functions are $h(\mu_{ijk})=\log(\mu_{ijk})$ and $h(\mu_{ijk})=\mu_{ijk}^{-2}$, and if the diffusion process results in a Gamma distribution for the RT, typical link functions are $h(\mu_{ijk})=\log(\mu_{ijk})$ and $h(\mu_{ijk})=\mu_{ijk}^{-1}$ [see \shortciteA{dunn2018generalized}, Chapter 11]. A key contribution of this work is establishing the connection between the parameters of the underlying cognitive process's diffusion, the covariates, and the fixed and random effects within the GLMM framework. This connection enables us to estimate the parameters of a diffusion process like Equation \eqref{eq:dif} for the underlying cognitive process associated with a given response by estimating the GLMM parameters.

To estimate the probability of choosing each response, we can utilize a frequency-based approach. Let $\mathcal{I}$ represent the set of all subscripts $(ijk)$. For each response $R_{l}$, where $l=1,2,\ldots,L$, we define $\mathcal{I}_{R_{l}}$ as the subset of $\mathcal{I}$ containing all trials with response $R_{l}$. The set of all responses in the experiment can then be expressed as the following disjoint union:
\[
\mathcal{I}=\bigcup\limits_{l=1}^{L}\mathcal{I}_{R_{l}},\]
For the RTs $Y=\{Y_{k}\}_{k=1}^m$, we define $\Omega_{R_{l}}=\{Y_{ijk}:(ijk)\in\mathcal{I}_{R_{l}}\}$ as the set of $Y|R_{l}$ results. Applying the Law of Large Numbers and Slutsky's theorem, we obtain:
\begin{equation}\label{eq:responsesprobabs}
\lim_{m\rightarrow\infty}\frac{1}{m}\sum_{i=1}^n\sum_{j=1}^{n_i}\sum_{k=1}^m\frac{{\rm card}(\Omega_{R_{l}}^{(i,j,k)})}{n_i}=\sum_{i=1}^n\sum_{j=1}^{n_i}\frac{P({R_{l}}^{(ij)})}{n_i},
\end{equation}
where convergence is in probability, and $R_{l}^{(ij)}$ denotes the event of responding $R_{l}$ in the $j$-th trial under the $i$-th stimulus level. Since we only consider individual-level random effects and maintain the same fixed effects for all $j$, the probability $P({R_{l}}^{(ij)})$  effectively depends only on $i$. Consequently, Equation \eqref{eq:responsesprobabs} simplifies to $\sum_{i=1}^nP({R_{l}}^{(i)})$. Alternatively, and  in a more structured modeling, we could employ regression models for the responses themselves, as explored in various studies [e.g., \shortciteA{van2007hierarchical}; \shortciteA{moscatelli2012modeling}; \shortciteA{molenaar2015bivariate}; \shortciteA{ranger2020modeling}].  
While the modeling of the responses themselves is not the focus of the present work, a more detailed discussion of such potential approaches is provided in Section \ref{sec:discussion}.

It is important to note that Equation \eqref{eq:responsesprobabs} relies on a large number of individuals in the experimental design for its validity. This requirement is often overlooked in practice. For instance, ``Experiment 1'' in \shortciteA{suarez2015dual} involved only 18 participants ($m=18$), each completing two sessions (``blocks'') of 96 trials.  Participants were divided into two groups: in the ``red left condition'' 50\% of participants responded to the color red with their left hand and green with their right, irrespective of the stimulus position on the screen; and in the ``red right condition'' the remaining 50\% responded to red with their right hand and green with their left. Stimuli were classified as ``congruent'' if the stimulus and assigned hand were on the same side, and ``incongruent'' otherwise.  This resulted in four stimuli (``green left'', ``red left'', ``green right'', ``red right'') and two conditions (``red left'', ``red right''). Combining stimuli and conditions yielded eight possible combinations ($n=8$) each representing a different level ($i=1,...,8$). Trials were indexed by $j=1,...,n_i$, with $\sum_{i=1}^8n_i=2\times96$. Due to the small sample size in this experiment, the consistency result in Equation \eqref{eq:responsesprobabs} is not applicable.

It is worth mentioning that incorporating multiple alternative responses can introduce additional temporal and error factors. These factors arise from the time required for an individual to decide how to express their desired response (e.g., which button to press). Our model does not explicitly account for these factors, highlighting an advantage of binary choice tasks over multiple-choice (MC) tasks. Additionally, we do not consider potential fatigue effects that may arise in longer testing sessions.

Parameter estimation methods for GLMMs are well-established and extensively documented in the literature [see, e.g., \shortciteA{molenberghs2002review}, \shortciteA{tuerlinckx2006statistical}, \shortciteA{stroup2012generalized}, \shortciteA{lee2018generalized}]. These methods are readily available in popular statistical software packages like R [see also \shortciteA{ronnegaard2010hglm}, \shortciteA{wang2022computation}]. Once we have estimated  a GLMM using real data (ideally achieving good fits with a large dataset), we can investigate several aspects relevant to the experimenter: (i)  stimuli and response differences aiming to analyze significant differences in RTs across various stimuli, levels, and response alternatives; (ii) response frequencies aiming to examine the frequency of different response alternatives, or the frequency of correct responses, in relation to different stimuli and their levels; and (iii) cognitive processing characteristics aiming to explore the underlying cognitive processing by establishing a theoretical link between the parameters of a mathematical model for cognitive processing and the parameters/covariates obtained from the GLMM applied to RTs conditioned on each response alternative. As mentioned above, we will focus mainly on aspects (i) and (iii), since the modeling of the responses themselves can be done using different approaches (see Section \ref{sec:discussion}).
We will illustrate (iii) with simple numerical examples.

\subsection{Parametrization and mapping}\label{sec:par-map}

We first examine the diffusion process in Equation \eqref{eq:dif} under the following setup: $\theta=0$, $\{e_t\}_{t\in\mathbb{R}_+}$ is a standard Brownian motion (that is, $e_0=0$ almost surely; for each $t\geq0$, $e_t\sim N(0,t)$ and has the same distribution as $e_{t+s}-e_s$ for any $s\geq0$; moreover, for every $0\leq s\leq t$, the increment $e_t-e_s$ is independent of $e_s$), and the initial condition is fixed at $X_0=a>0$. Under this specification, the hitting time of $X_t$ at 0 follows an inverse Gaussian (IG) distribution with mean $a/\nu$ and variance $a/\nu^3$ (equivalently, in the mean–shape parametrization, the distribution of this random variable is $IG(a/\nu,a^2)$). In our GLMM setting, by defining a suitable alternative response, the unconditional expectation for trial $j$ at stimulus level $i$ is then given by
\begin{equation}\label{eq:mean}
\mu_{ij}=\int E(Y_{ijk}|c)g(c)dc=E(Y_{ijk}).
\end{equation}
(Here, we use the term “unconditional” to refer to integration over the random effects; this should not be confused with removing conditioning on the responses, whose conditioning will always remain). Furthermore, the unconditional variance for trial $j$ at stimulus level $i$ is:
\begin{equation}\label{eq:var}
\sigma^2_{ij}=\int V(Y_{ijk}|c)g(c)dc+\int E(Y_{ijk}|c)^2g(c)dc-\mu_{ij}^2=V(Y_{ijk}).
\end{equation}
Therefore, if we consider Equation \eqref{eq:dif} as representing the underlying cognitive process of a typical individual responding to such an alternative response, under trial type $j$ and stimulus level $i$, respectively, we can rewrite it as:
\[
X^{(ij)}=a_{ij}-\nu_{ij}t+e^{(ij)}_{t},
\]
where the $e^{(ij)}_{\cdot}$'s, with $e^{(ij)}_{\cdot}=\{e^{(ij)}_{t}\}_{t\in\mathbb{R}_+}$, are IID Brownian motions. Conditionally on the random effect (i.e.\ for a typical individual), this hitting time is exactly IG with mean $a_{ij}/\nu_{ij}$ and variance $a_{ij}/\nu_{ij}^3$. Integrating over the random effect, the marginal law of $Y_{ijk}$ is a \emph{mixture} of such IG distributions and is therefore not exactly IG; we reconstruct the cognitive process by matching the first two \emph{marginal} moments \eqref{eq:mean}--\eqref{eq:var} to the IG moment relations, $\mu_{ij}=a_{ij}/\nu_{ij}$ and $\sigma^2_{ij}=a_{ij}/\nu_{ij}^3$. This marginal matching is a \textit{moment-based approximation}; an exact alternative is to reconstruct conditionally at $c=0$, where $Y_{ijk}\mid c=0$ is exactly IG.

So, we can estimate the underlying cognitive process from the GLMM \emph{marginal} moment estimates $\hat{\mu}_{ij}$ and $\hat{\sigma}^2_{ij}$ for each alternative response---whose family-specific expressions under the log link and normal random effects are derived in Section~\ref{sec:experiment}---by inverting the IG moment relations: $\hat{\nu}_{ij}=\sqrt{\hat{\mu}_{ij}/\hat{\sigma}^2_{ij}}$ and $\hat{a}_{ij}=\hat{\mu}_{ij}\hat{\nu}_{ij}$. This estimation provides a distributional representation of the corresponding cognitive mechanism leading to the observed response.

\textit{Remark.} Since only individual-level random effects are considered and the fixed effects remain constant across all trials $j$, Equations \eqref{eq:mean} and \eqref{eq:var} become independent of $j$. This framework can, however, be extended to include random effects within individuals, as discussed in Section \ref{sec:discussion}.

\begin{figure}[htbp]
\centering
\includegraphics[width=0.40\textwidth]{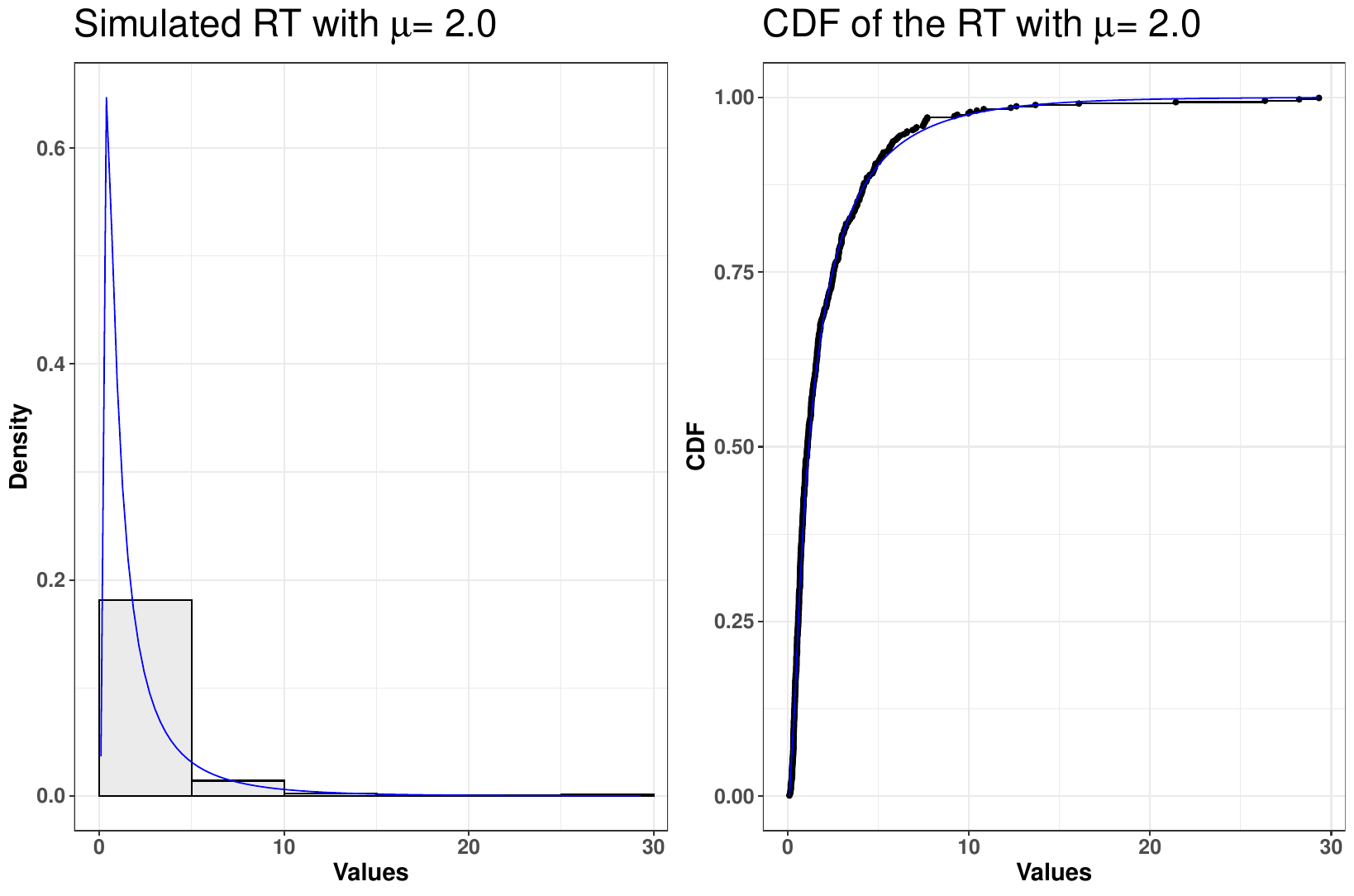}
\includegraphics[width=0.40\textwidth]{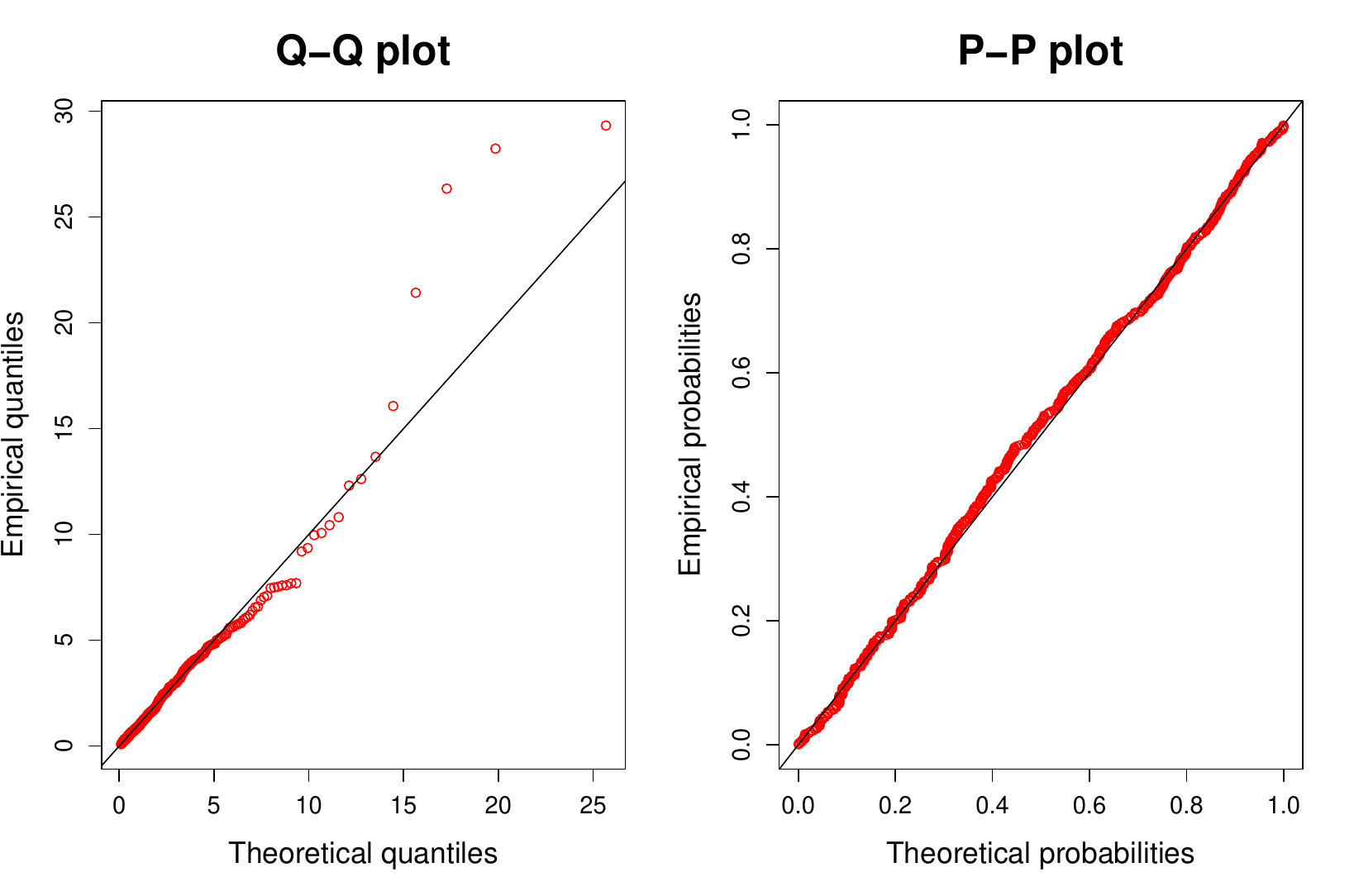}
\includegraphics[width=0.40\textwidth]{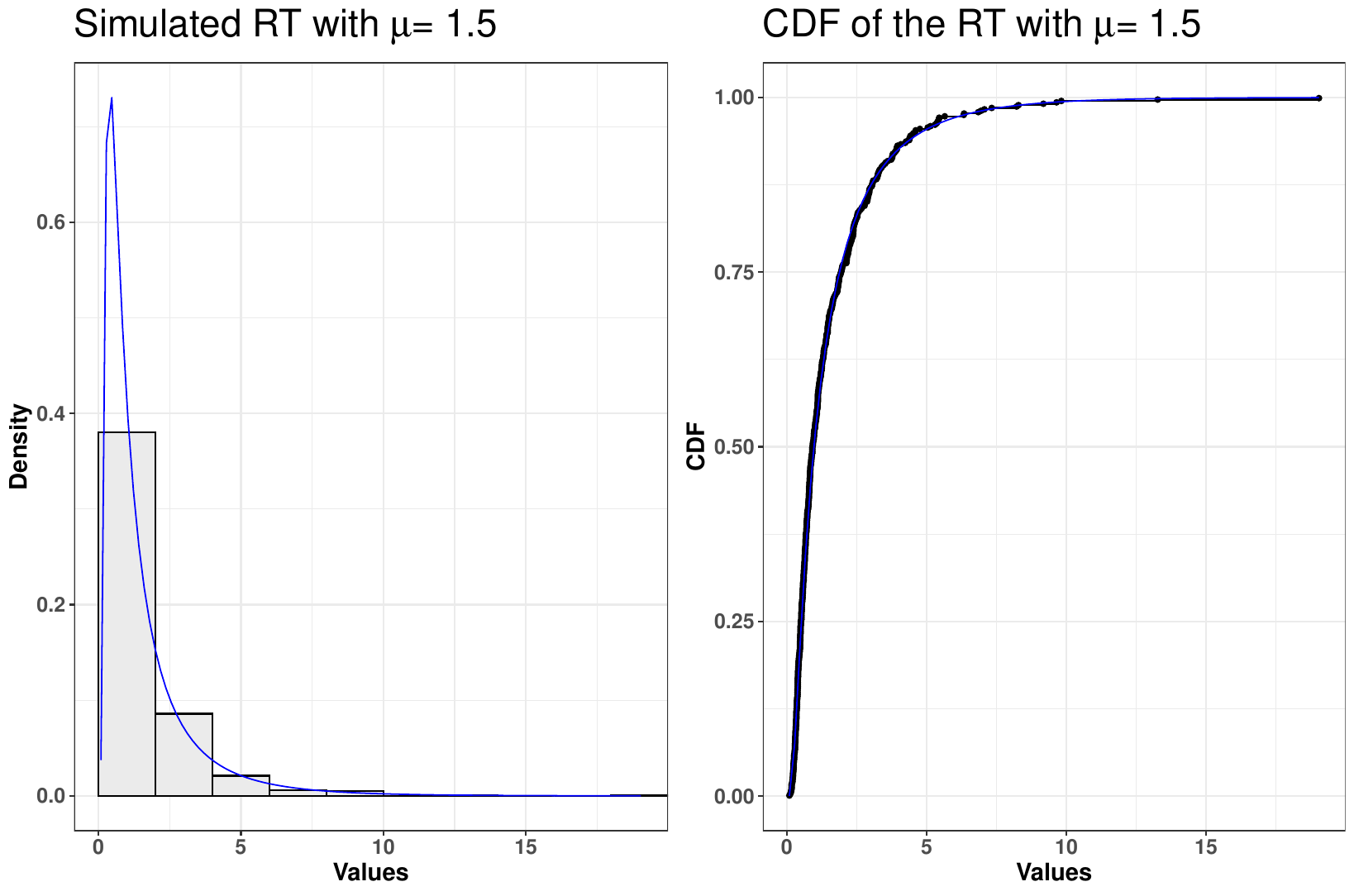}
\includegraphics[width=0.40\textwidth]{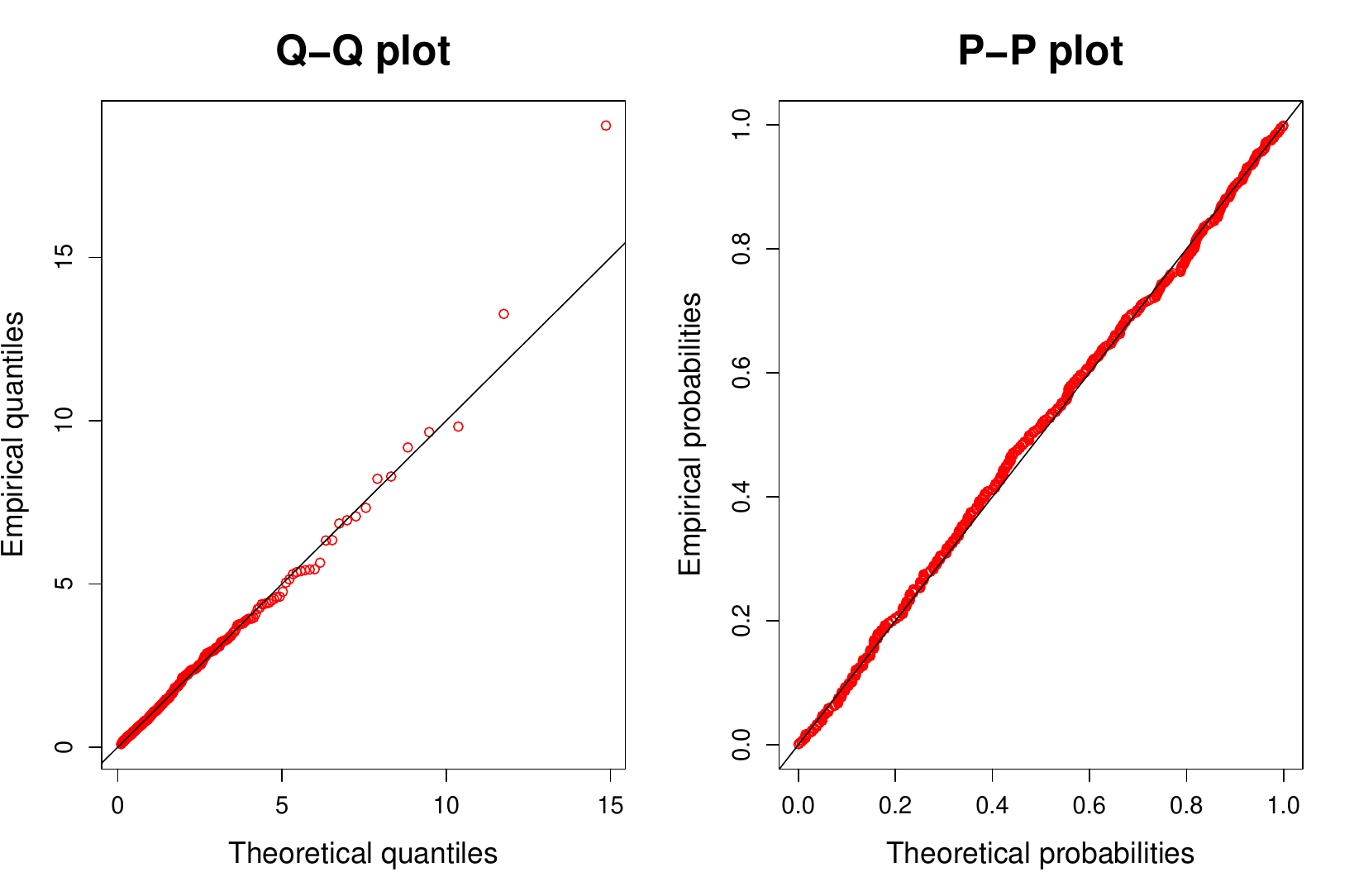}
\includegraphics[width=0.40\textwidth]{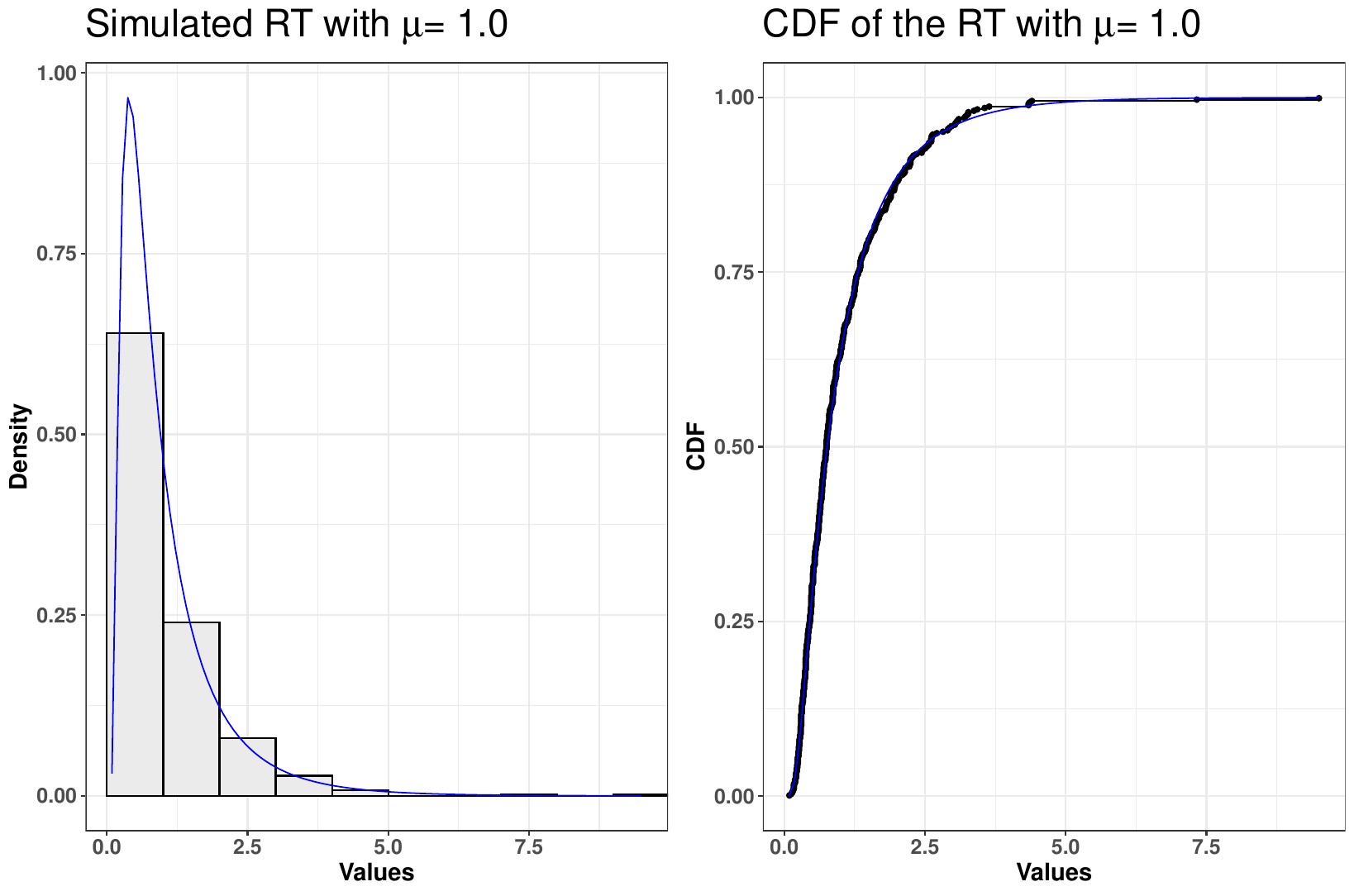}
\includegraphics[width=0.40\textwidth]{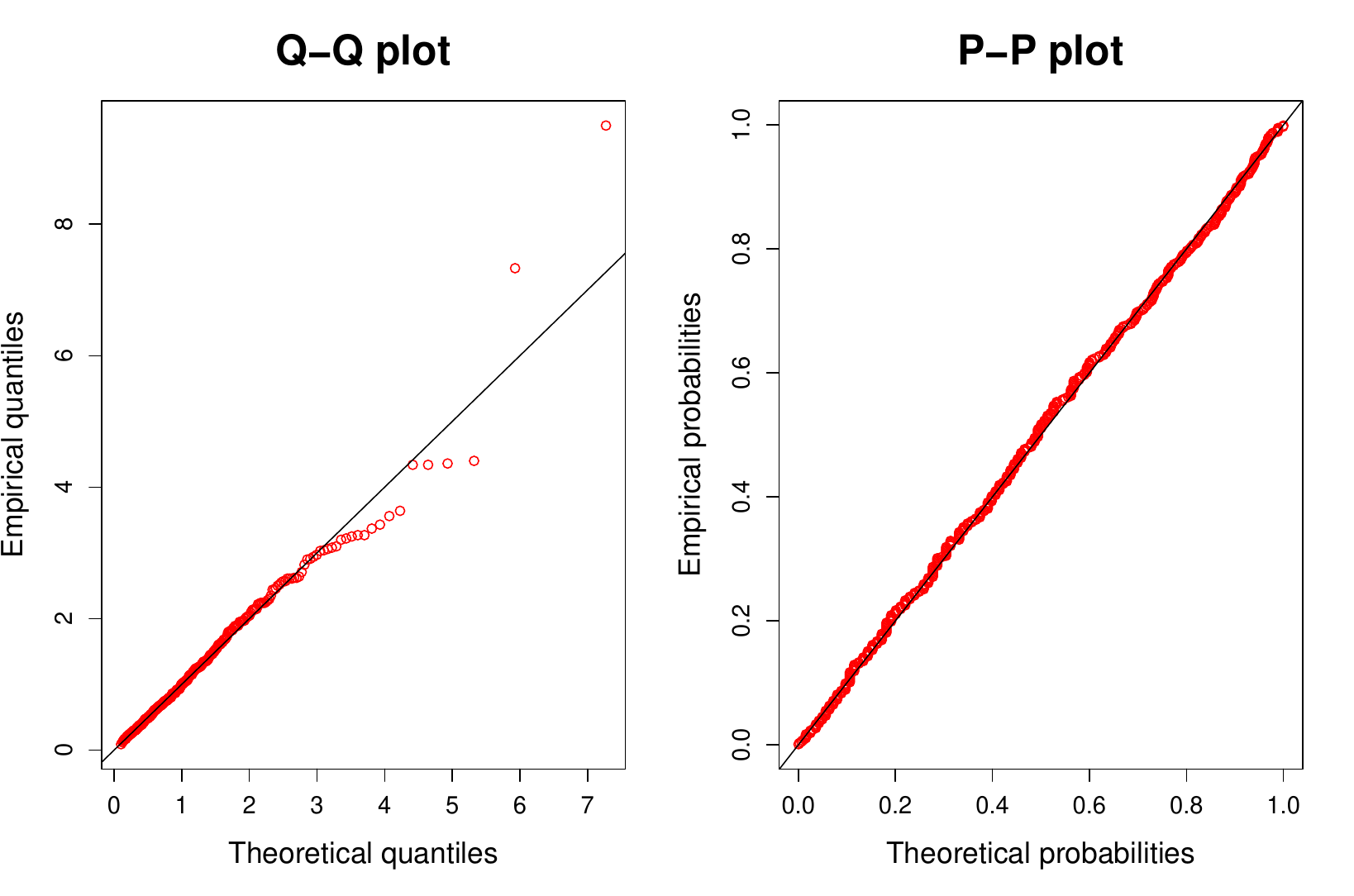}
\includegraphics[width=0.40\textwidth]{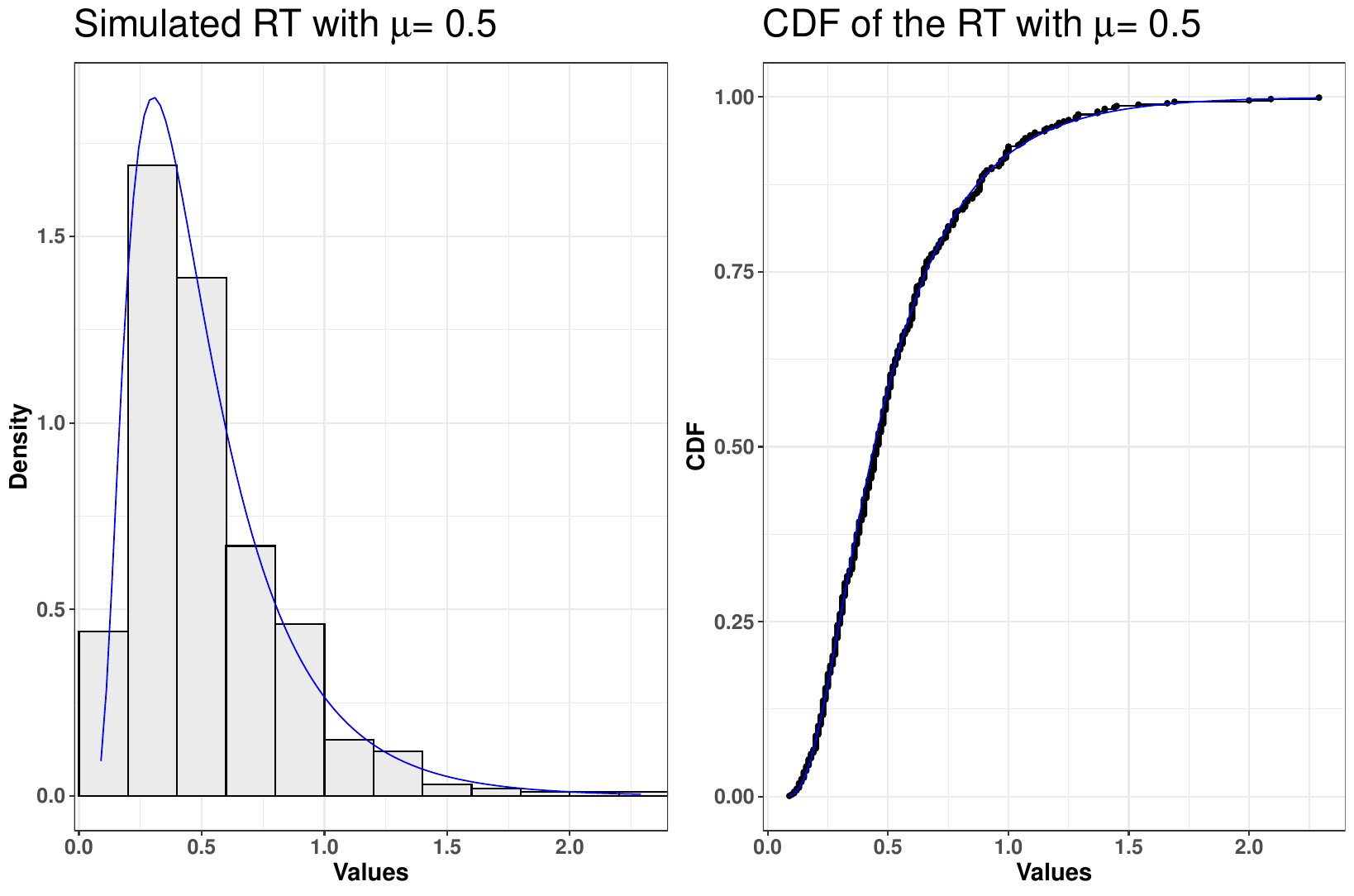}
\includegraphics[width=0.40\textwidth]{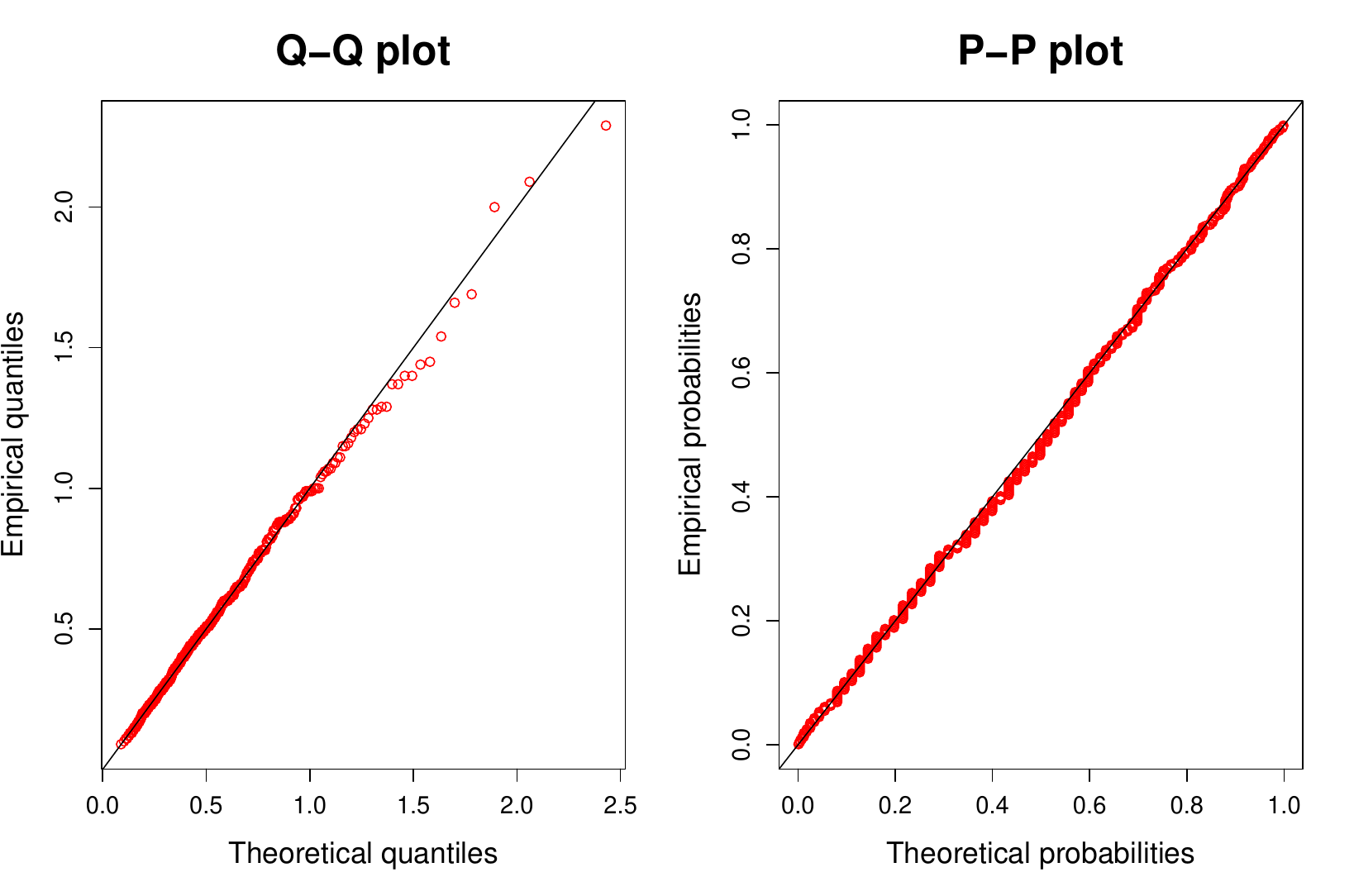}
\caption{Plots displaying the goodness-of-fit assessment for simulated IG data. Each row corresponds to a different simulated dataset with a specific mean ($\mu=2$, $\mu=1,5$, $\mu=1$ and $\mu=0.5$, from top to bottom). In all cases, the scale parameter ($\phi$) was set to 1. From left to right, each row shows the estimated PDF via a histogram of the simulated data with the theoretical IG PDF overlaid in blue; estimated CDF showing the empirical CDF of the simulated data with the theoretical IG CDF overlaid in blue; Quantile-Quantile (Q-Q) Plot comparing the quantiles of the simulated data to the quantiles of the theoretical IG distribution; and Probability-Probability (P-P) Plot comparing the cumulative probabilities of the simulated data to the cumulative probabilities of the theoretical IG distribution. For each case, 1000 random variables $Y_1$, \ldots, $Y_{1000}$ were simulated, and the recursive scheme was replicated 500 times to obtain the corresponding RTs ($T^*_1$,\ldots, $T^*_{500}$).}\label{fig:IGsim}
\end{figure}

To illustrate this concept in a simplified setting, we simulate independent and identically distributed (IID) inverse Gaussian (IG) responses, denoted as ${Y_u}_u$, with mean parameter $\mu$ and shape parameter fixed at $1$. Hence, $E(Y_u)=\mu$ and $V(Y_u)=\mu^3$. In this context, the mean--variance relationship in \eqref{eq:mean-var}, here given by $V(Y_u)=\phi E(Y_u)^{\lambda}$, can be expressed by setting $\lambda=3$ and $\phi=1$. We generate four datasets, each consisting of 1000 observations, with $\mu=2$, $\mu=1.5$, $\mu=1$, and $\mu=0.5$, respectively. For each dataset, we estimate the corresponding values of $\mu$ and $\phi$. We then simulate the respective response times based on Equations \eqref{eq:dif}-\eqref{stopping} using the following scheme:

\begin{enumerate}
\item Fix $X_0=\sqrt{1/\hat{\phi}}$ and $\delta=0.01$.
\item Iterate recursively
\[
X_{(t)}=X_{(t-1)}-\frac{1}{\hat{\mu}}\sqrt{\frac{1}{\hat{\phi}}}\delta+\sqrt{\delta}\epsilon_{(t)},
\]
with $t=1,2,\ldots$, and where the $\epsilon_{(t)}$'s are IID standard normal distributed.
\item The above process continues until the first $t=t^*$ such that $X_{(t^*)}\leq0$.
\item The simulated RT is given by $T^*=t^*\delta$.
\item We replicate this $M$ times, obtaining a random sample of RTs $T^*_1$, $T^*_2$,\ldots, $T^*_M$.
\item Finally, we assess the goodness-of-fit of these simulated RT datasets against the theoretical IG distributions they were originally generated from (those with $\mu=2$, $\mu=1,5$, $\mu=1$ and $\mu=0.5$, and with $\phi=1$ in each case). The results of this goodness-of-fit evaluation are presented in Figure \ref{fig:IGsim}.
\end{enumerate}

Now consider a diffusion process of the form given in Equation \eqref{eq:dif}, with $\theta = 0$ and $\{e_t\}_{t \in \mathbb{R}_+}$ a standard Brownian motion, along with its associated stopping time defined in \eqref{stopping}. For the Gamma model, we examine the following inverse problem: if this stopping time follows a Gamma distribution with shape parameter $\alpha$ and scale parameter $\beta$, then there exists a diffusion process as in Equation \eqref{eq:dif} such that $\nu = \sqrt{\beta/2}^{-1}$ and $X_0$ is distributed as the sum of two independent Gamma random variables, each with shape parameter $\alpha$ and scale parameter $\sqrt{\beta/2}$. In this construction, the first-hitting time of $X_t$ to zero follows a Gamma distribution with shape $\alpha$ and scale $\beta$ (see \shortciteA{jackson2009randomization} for the original extended result; for a reaction time context, see \shortciteA{tejo2019theoretical}, where this result can be applied with $\theta > 0$, in which case the corresponding stopping time follows a ``shifted'' Gamma distribution).

In the context of our GLMM, and focusing on a given alternative response, the unconditional expectation and variance for trial $j$ under stimulus level $i$ are provided by Equations \eqref{eq:mean} and \eqref{eq:var}, respectively. Accordingly, if we interpret Equation \eqref{eq:dif} as characterizing the latent cognitive process of a typical individual producing such an alternative response, under trial type $j$ and stimulus level $i$, it can be expressed as:
\[
X^{(ij)}=X_0^{(ij)}-\sqrt{\beta_{ij}/2}^{-1}t+e^{(ij)}_{t},
\]
where the $X_0^{(ij)}$'s are independent, each distributed as the sum of two independent Gamma random variables with common shape and scale parameters $\alpha_{ij}$ and $\sqrt{\beta_{ij}/2}$, respectively, and the $e^{(ij)}_{\cdot}$'s are again IID Brownian motions. As in the IG case, conditionally on the random effect the hitting time is exactly Gamma with mean $\alpha_{ij}\beta_{ij}$ and variance $\alpha_{ij}\beta_{ij}^2$, whereas the marginal law of $Y_{ijk}$ is a \emph{mixture} of Gamma distributions and hence not exactly Gamma; we again reconstruct by matching the first two \emph{marginal} moments \eqref{eq:mean}--\eqref{eq:var} to the Gamma moment relations, $\mu_{ij}=\alpha_{ij}\beta_{ij}$ and $\sigma^2_{ij}=\alpha_{ij}\beta_{ij}^2$ (a moment-based approximation marginally, exact conditionally at $c=0$).

Again, we estimate the underlying cognitive process from the GLMM \emph{marginal} moment estimates $\hat{\mu}_{ij}$ and $\hat{\sigma}^2_{ij}$ (family-specific expressions in Section~\ref{sec:experiment}) by inverting the Gamma moment relations: $\hat{\alpha}_{ij}=\hat{\mu}_{ij}^2/\hat{\sigma}^2_{ij}$ and $\hat{\beta}_{ij}=\hat{\sigma}^2_{ij}/\hat{\mu}_{ij}$. This estimation provides a distributional representation of the corresponding cognitive mechanism leading to the observed response.

\textit{Remark.} As noted in the previous remark, because only individual-level random effects are considered and the fixed effects remain constant across all trials $j$, Equations \eqref{eq:mean} and \eqref{eq:var} become independent of $j$ (see Section \ref{sec:discussion} for potential extensions).

\begin{figure}[htbp]
\centering
\includegraphics[width=0.40\textwidth]{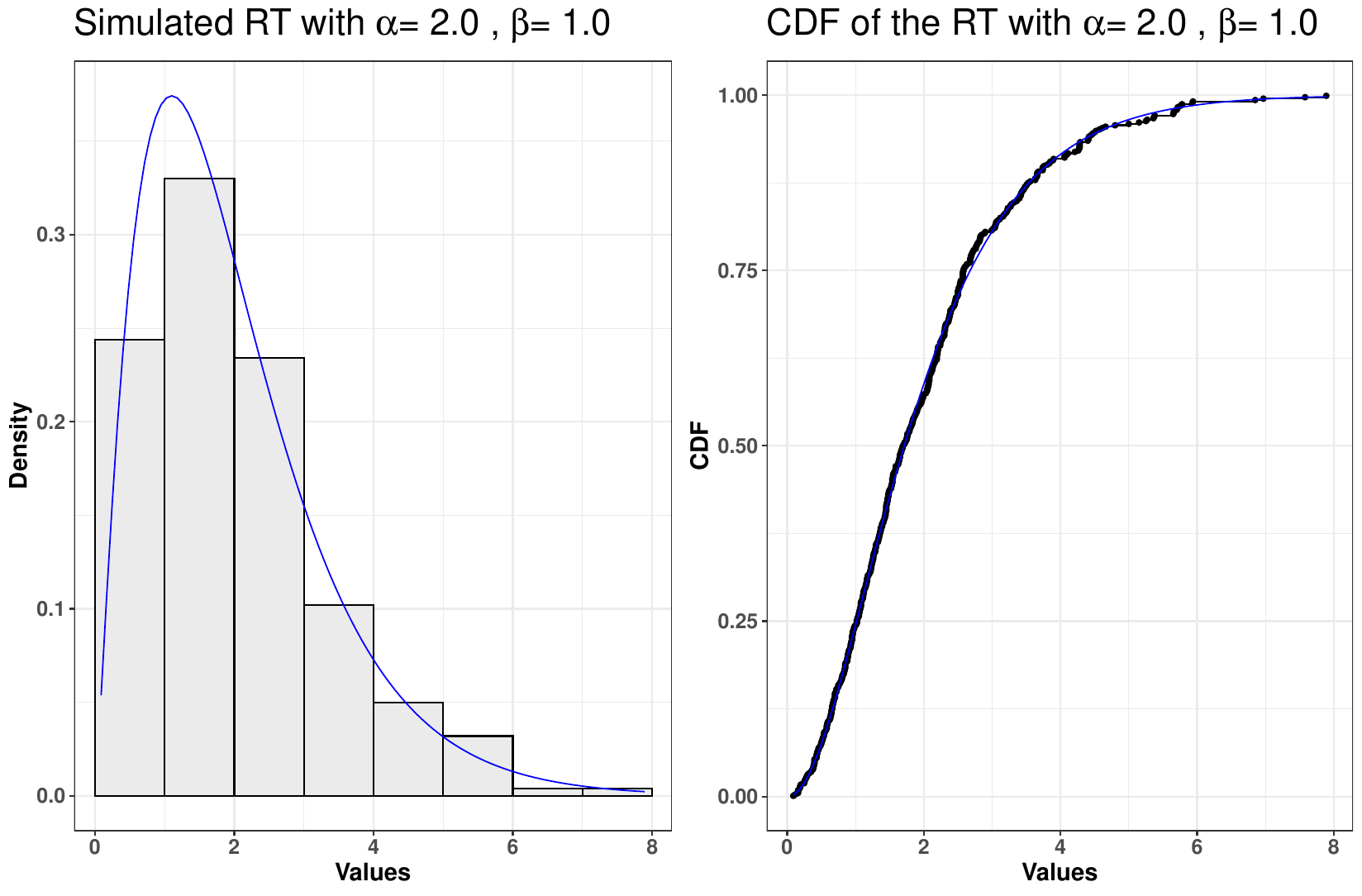}
\includegraphics[width=0.40\textwidth]{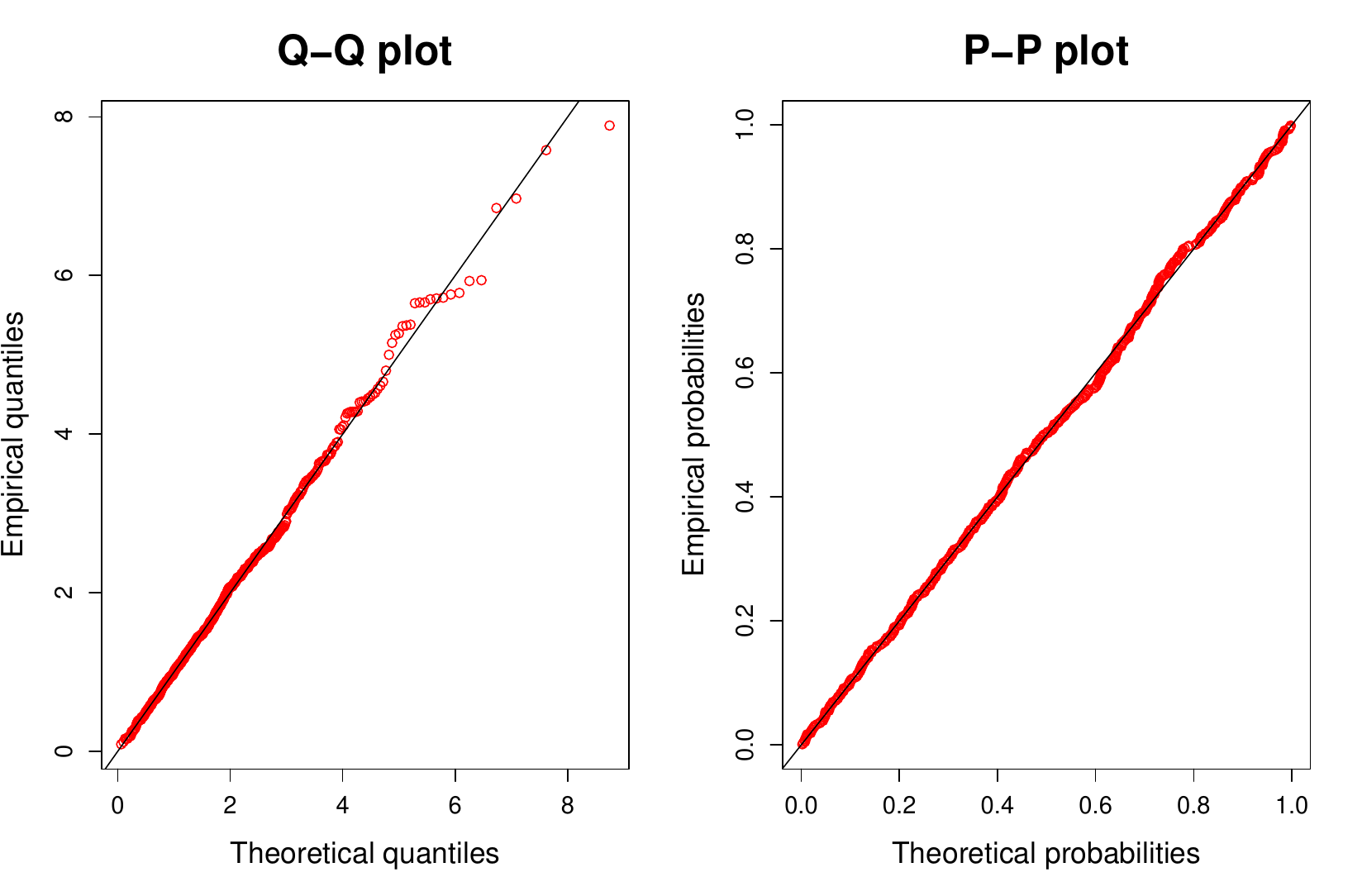}
\includegraphics[width=0.40\textwidth]{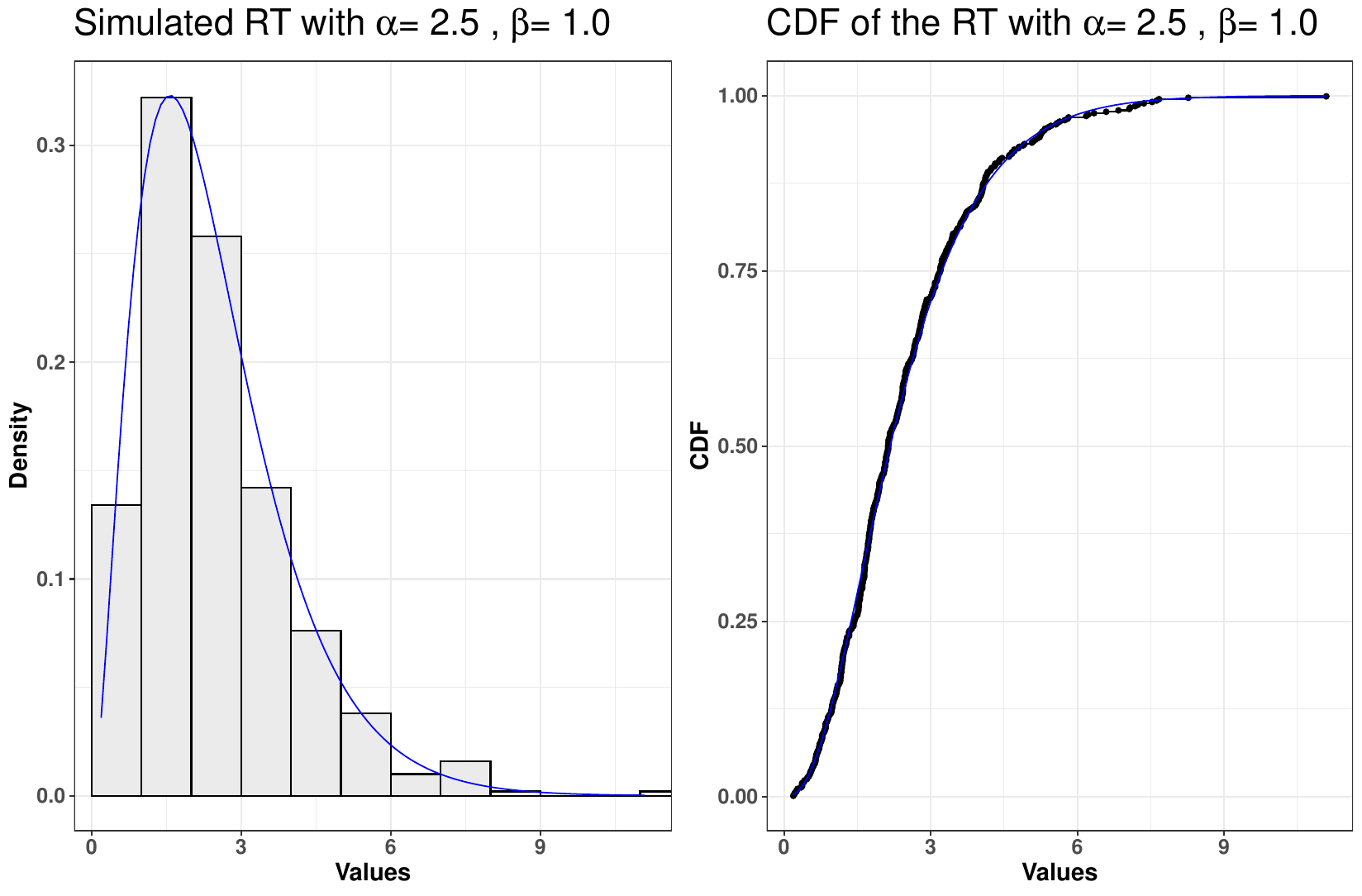}
\includegraphics[width=0.40\textwidth]{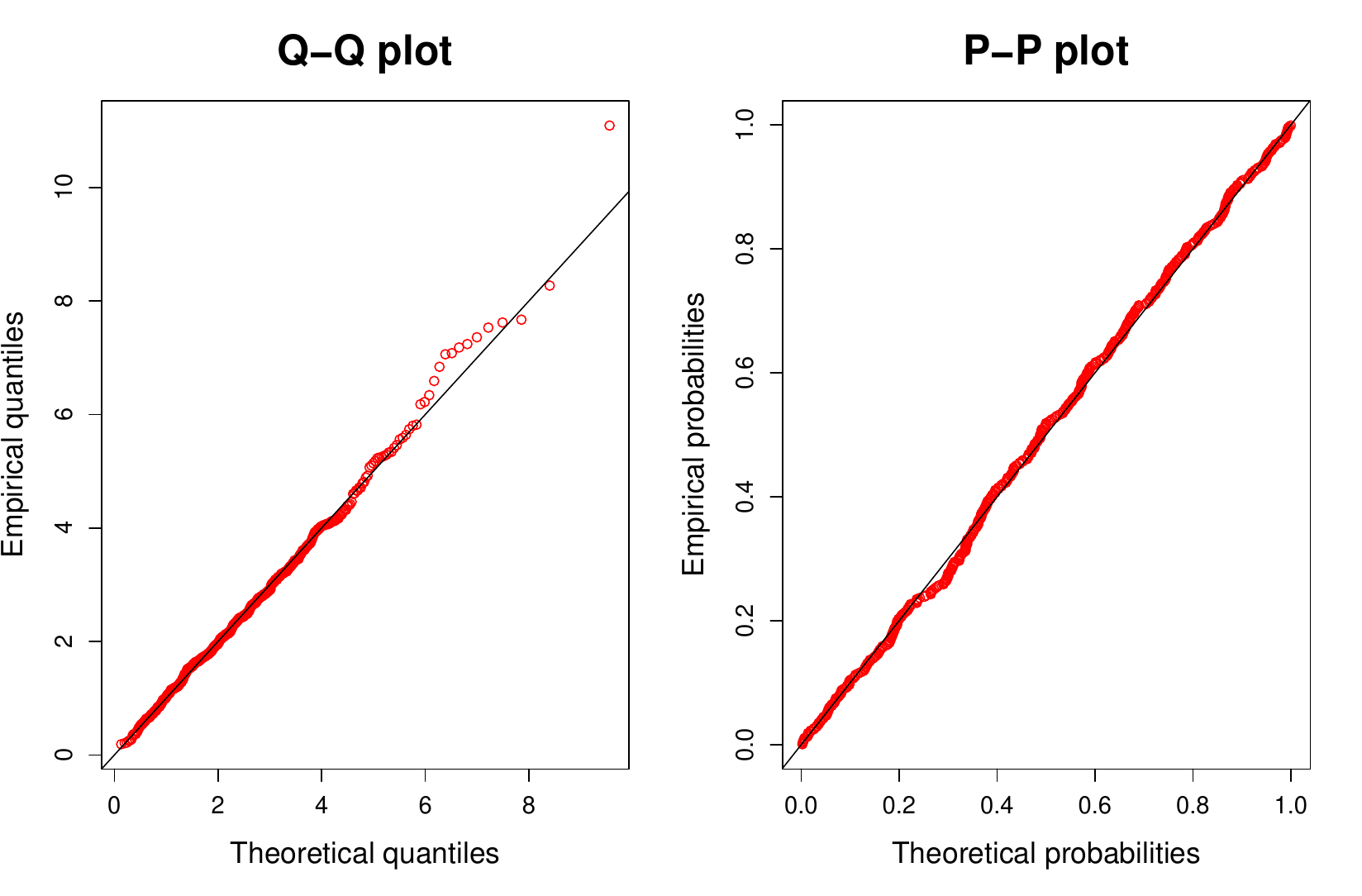}
\includegraphics[width=0.40\textwidth]{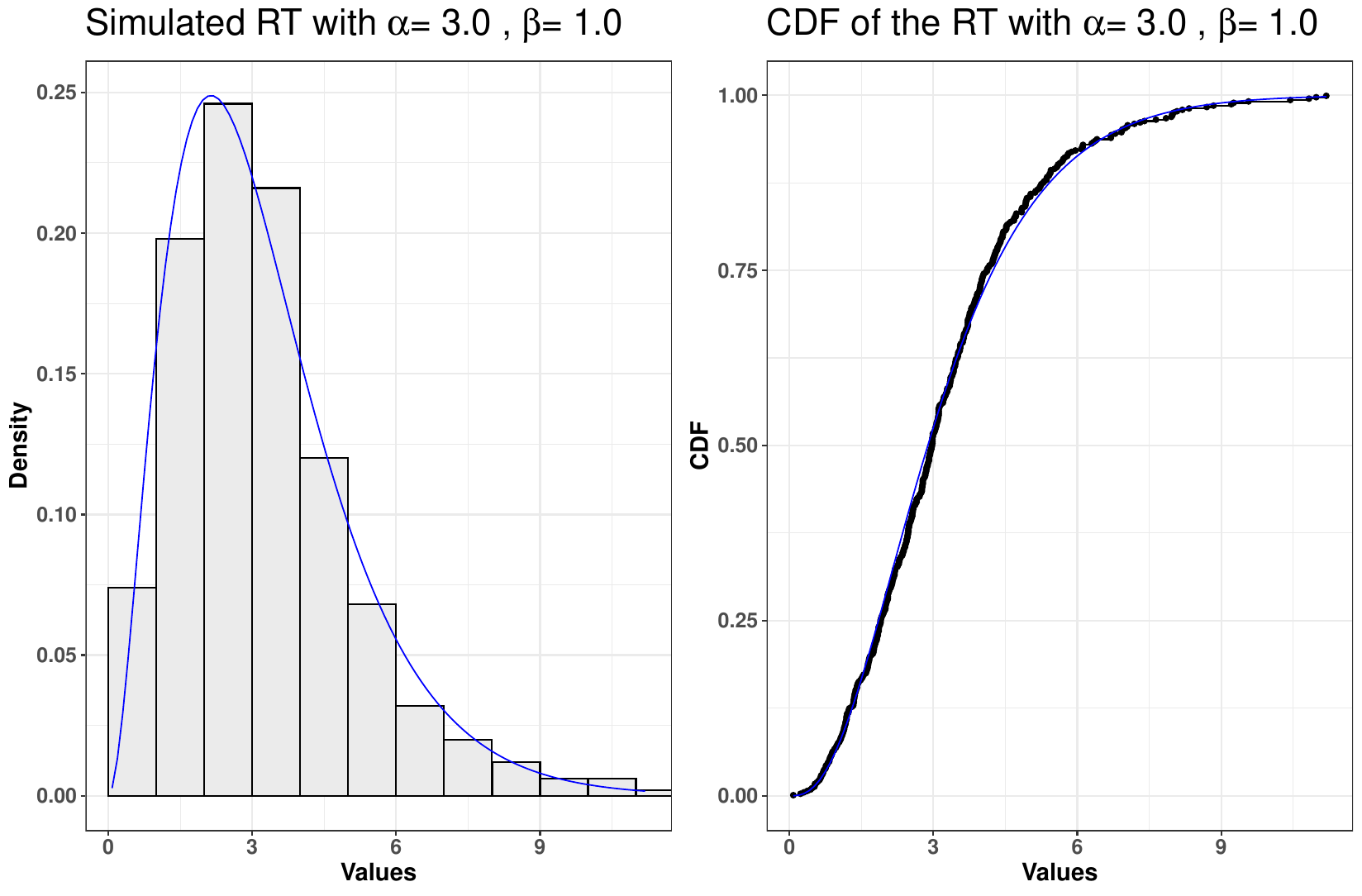}
\includegraphics[width=0.40\textwidth]{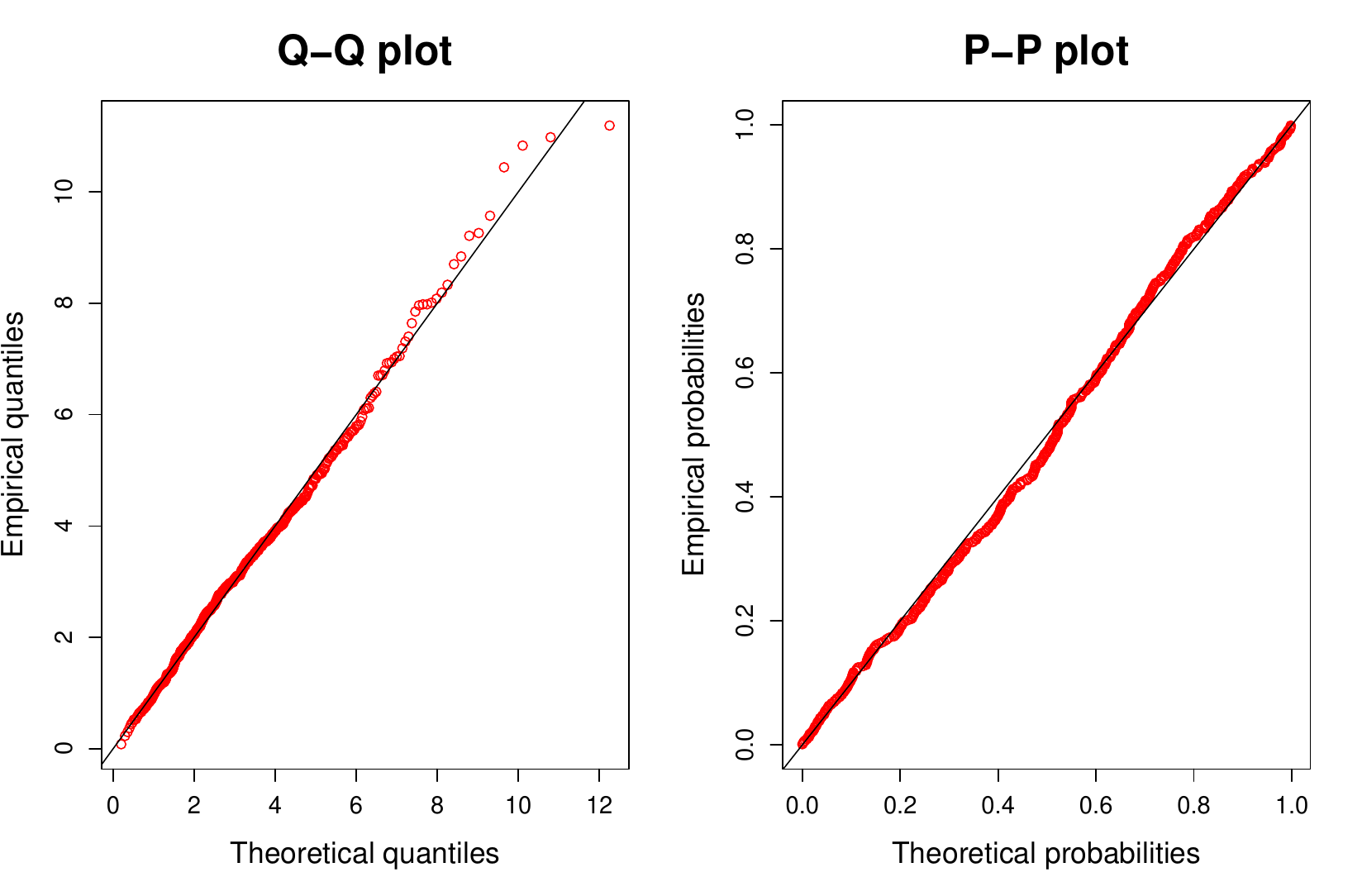}
\includegraphics[width=0.40\textwidth]{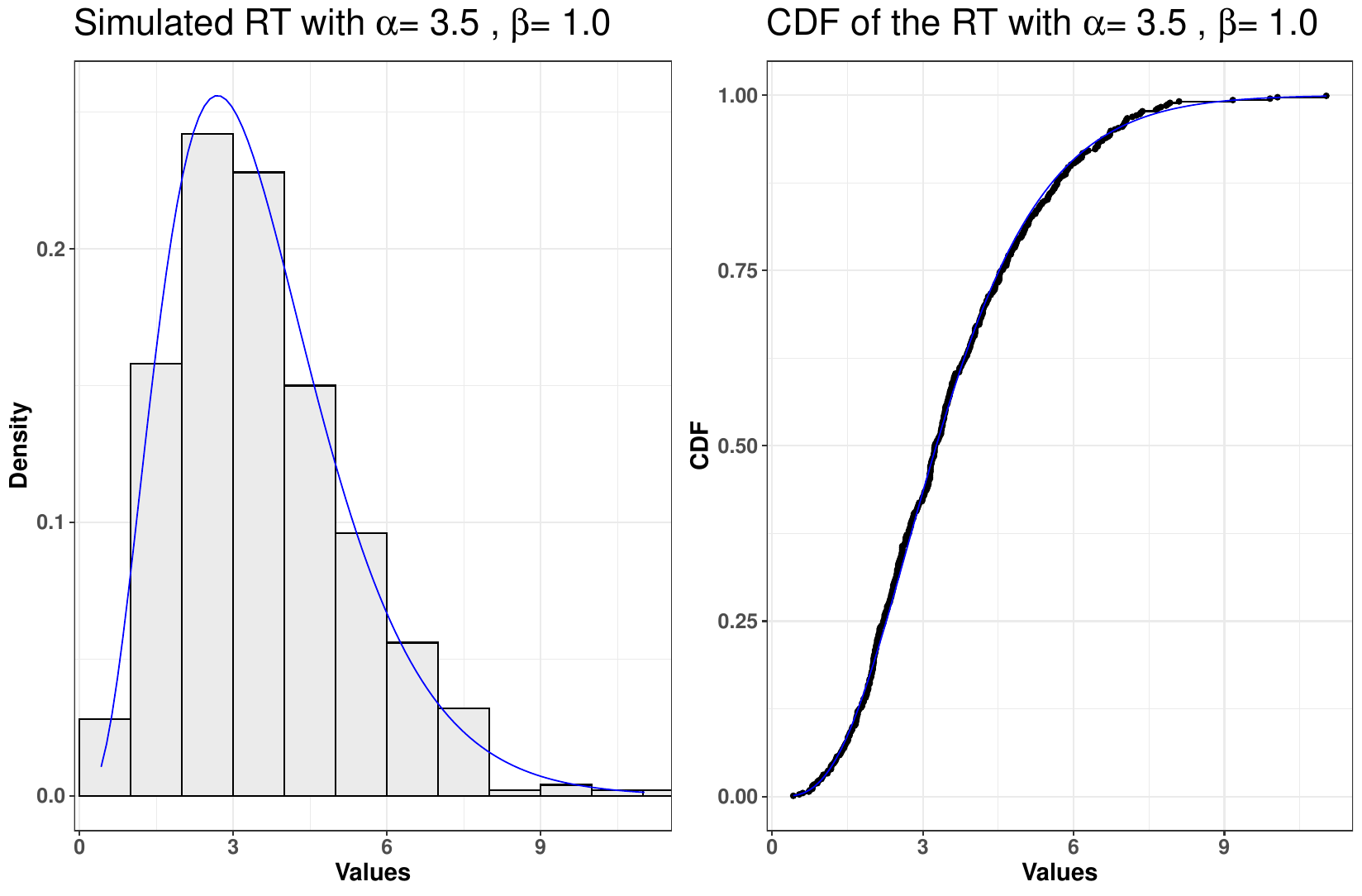}
\includegraphics[width=0.40\textwidth]{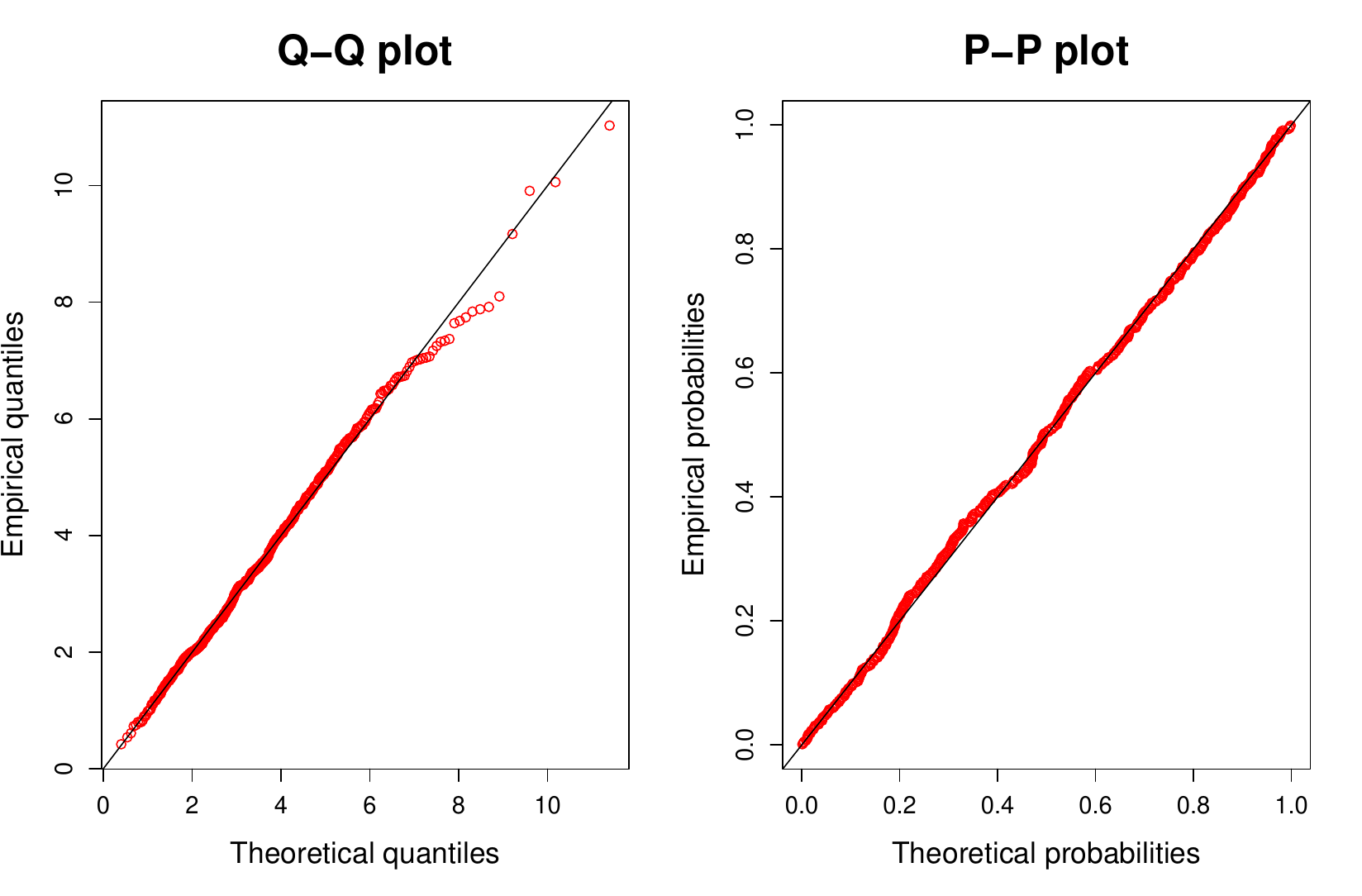}
\caption{Plots displaying the goodness-of-fit assessment for simulated Gamma data. Each row corresponds to a different shape parameter value ($\alpha=2$, $\alpha=2.5$, $\alpha=3$ and $\alpha=3.5$, from top to bottom). In all cases, the scale parameter ($\beta$) was set to 1. From left to right, each row shows the estimated PDF via a histogram of the simulated data with the theoretical Gamma PDF overlaid in blue; estimated CDF showing the empirical CDF of the simulated data with the theoretical Gamma CDF overlaid in blue; Quantile-Quantile (Q-Q) Plot comparing the quantiles of the simulated data to the quantiles of the theoretical Gamma distribution; and Probability-Probability (P-P) Plot comparing the cumulative probabilities of the simulated data to the cumulative probabilities of the theoretical Gamma distribution. For each case, 1000 random variables $Y_1$, \ldots, $Y_{1000}$ were simulated, and the recursive scheme was replicated 500 times with $\delta=0.01$ to obtain the corresponding RTs ($T^*_1$,\ldots, $T^*_{500}$).}\label{fig:GMsim}
\end{figure}

We can develop an algorithm analogous to the simplified scenario above to simulate response times following a Gamma distribution and to recover the underlying cognitive process described by Equation \eqref{eq:dif}. The procedure is as follows: (i) Generate independent and identically distributed (IID) Gamma responses, denoted ${Y_u}_u$, with shape parameter $\alpha$ and scale parameter $\beta$. Accordingly, $E(Y_u) = \alpha\beta = \mu$ and $V(Y_u) = \alpha\beta^2 = (1/\alpha)\mu^2$. In this case, the mean--variance relationship in \eqref{eq:mean-var}, here given by $V(Y_u) = \phi E(Y_u)^\lambda$, can be expressed by setting $\lambda = 2$ and $\phi = 1/\alpha$. (ii) Estimate the parameters $\alpha$ and $\beta$ from the generated datasets. (iii) Simulate the corresponding response times based on Equations \eqref{eq:dif}--\eqref{stopping} using the following scheme:

\begin{enumerate}
\item Fix some $\delta$ small, and consider $\theta=0$. The initial value $X_0$ is taken from the sum of two independent Gamma with common shape and scale $\hat{\alpha}$ and $\sqrt{\hat{\beta}/2}$, respectively.

\item Iterate recursively
\[
X_{(t)}=X_{(t-1)}-\sqrt{\hat{\beta}/2}^{-1}\delta+\sqrt{\delta}\epsilon_{(t)},
\]
with $t=1,2,\ldots$, and where the $\epsilon_{(t)}$'s are IID standard normal distributed.

\item The above process continues until the first $t=t^*$ such that $X_{(t^*)}\leq0$.

\item The simulated RT is given by $T^*=t^*\delta$.

\item We replicate this $M$ times (independently), obtaining a random sample of RTs $T^*_1$, $T^*_2$, \ldots, $T^*_M$.

\item We finally verify the goodness-of-fit  of such RT sets with respect to the theoretical distribution for each data set originally simulated (See Fig. \ref{fig:GMsim}).

\end{enumerate}

We conclude this Section with the following important Remarks.

\textit{Remarks:} 

(i) Theoretical models, despite their simplicity, make it relatively straightforward to estimate the parameter $\theta$ in Equation \eqref{eq:dif} [\shortciteA{tejo2019theoretical}]. In contrast, the hierarchical specification in Equation \eqref{eq:hiestruc}—where we incorporate all trials for each stimulus level and condition them on each possible response—poses a substantial difficulty when attempting to introduce and then estimate $\theta$. In this framework, we would in fact have multiple parameters, $\theta_{ij}$, one for each trial $j$ at stimulus level $i$, and each also depending on the corresponding response alternative. In this work, we proceed with the reconstruction by fixing $\theta=0$, acknowledging that this choice may introduce some bias. Specifically, the RT $Y$, defined via an equation such as \eqref{stopping}, can be written as $Y=\tilde{Y}-\theta$, where $\tilde{Y}$ represents the entire non-decision/decision-related process. This is the process that we are effectively modeling when we take $\theta=0$. The explicit inclusion of this parameter within the GLMM framework is left for future work.

(ii) In simple diffusion models, it is common to introduce an extra parameter into the latent cognitive process to represent a diffusion coefficient. For instance, one might write
$X_t = X_0 - \nu t + \varsigma e_t$, where $\varsigma > 0$ is such a parameter capturing the magnitude of the Brownian noise $e_\cdot$. However, even though this parameter cannot be identified from the data for estimation, we should remember that the latent cognitive process is an “abstraction”: a schematic representation that cannot be directly observed in concrete measurement units. Consequently, allowing a diffusion coefficient $\varsigma$ different from 1 merely rescales the parameters of the diffusion model (i.e., we can equivalently write $X_t^* = X_0^* - \nu^* t + e_t$, where $X_t^* = X_t / \varsigma$ and $\nu^* = \nu / \varsigma$), while preserving the same distribution for the resulting response times (RTs).

The following section details the application of our methodology to ``Experiment 1'' documented in \shortciteA{marmolejo2020your}.

%%%%%%%%%%%%%%%%%%%%%%%%%%%%%%%%%%%%%%%%%%%%%%%%%5
\section{Experiment and statistical data analysis}\label{sec:experiment}

\textit{Description of the experiment and database.} Participants were instructed to identify a facial expression as ``sad'' or ``happy'' as quickly and accurately as possible, either while holding a pen in their teeth or with no pen. Each participant completed 154 trials, structured as $11$ stimuli $\times\, 2$ experimental conditions (pen-in-teeth and no-pen) $\times\, 7$ repetitions; thus each combined stimulus--condition level was observed seven times per participant [for a comprehensive description of the experiment, see \shortciteA{marmolejo2020your}]. Reaction times were analyzed in milliseconds (the raw values, recorded in seconds, were multiplied by $1000$). No trimming, winsorization or truncation was applied---all recorded trials were retained, including a few very long responses (up to ${\sim}34$ s). The analyzed dataset comprises $17{,}683$ trials: two participants have incomplete sessions (one with only $54$ trials), so the total is slightly below the nominal $116\times 154=17{,}864$, and the GLMMs use all available trials.

The 11 stimuli ranged from a ``fully sad'' facial expression (designated as ``stimulus 0'') to a ``fully happy'' one (designated as ``stimulus 10''). Given the two experimental conditions (pen-in-teeth and no-pen), we treated the combined levels of all stimulus types as a set of 22 unique levels.  These levels were indexed as follows: $i=1$ representing ``stimulus 0 + pen-in-teeth'', i=2 representing ``stimulus 1 + pen-in-teeth'', and so on, up to $i=11$ for ``stimulus 10 + pen-in-teeth''. The indexing continued with $i=12$ for ``stimulus 0 + no-pen'' and concluded with $i=22$ for ``stimulus 10 + no-pen''.

The study involved 116 participants $(k=1,...,116)$, each exposed to every combined stimulus level seven times across seven ``blocks''; thus, $j=1,...,7$ represents the block number. Participants provided binary responses, categorized as $R_1$: ``sad'' or $R_2$: ``happy''.

\begin{figure}[htbp]
\centering
\resizebox{6cm}{!}{\includegraphics{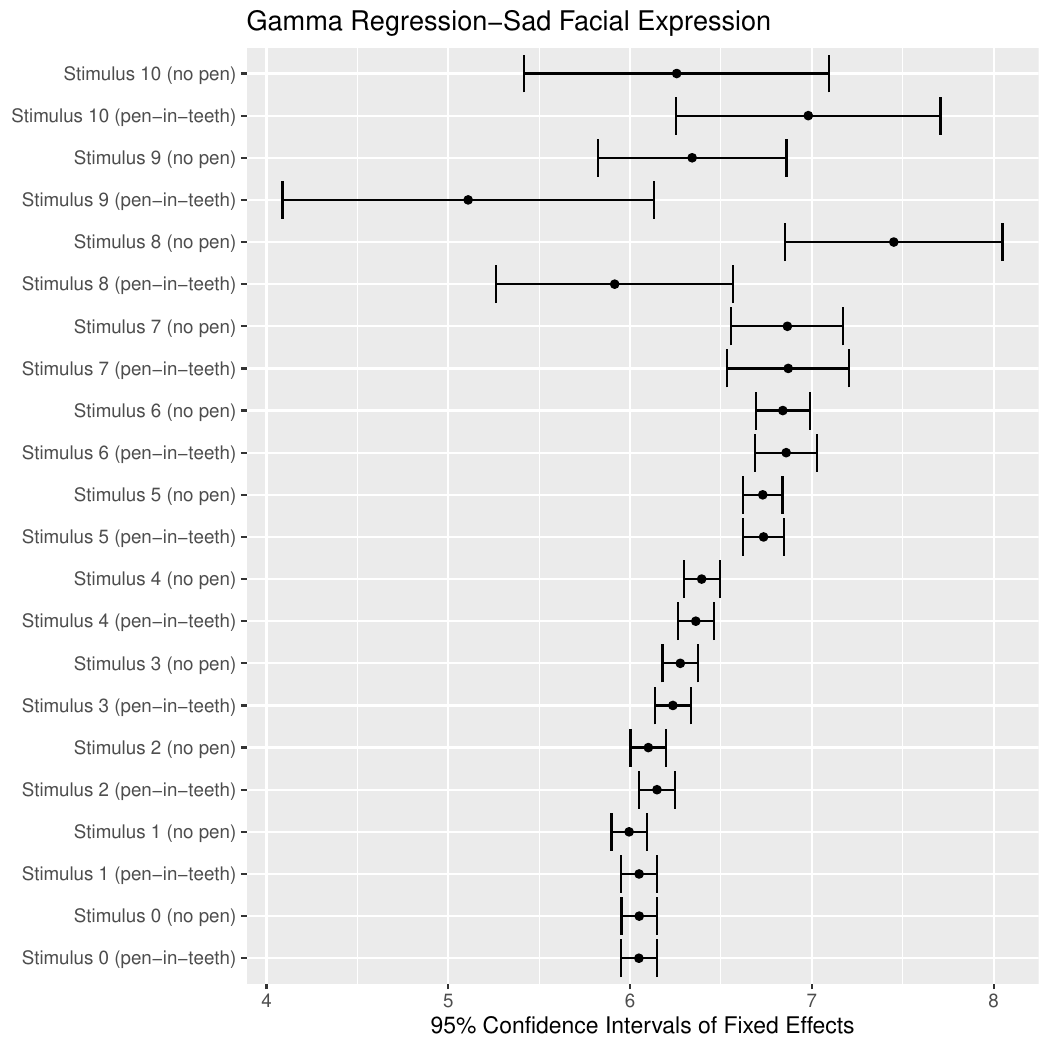}}
\resizebox{6cm}{!}{\includegraphics{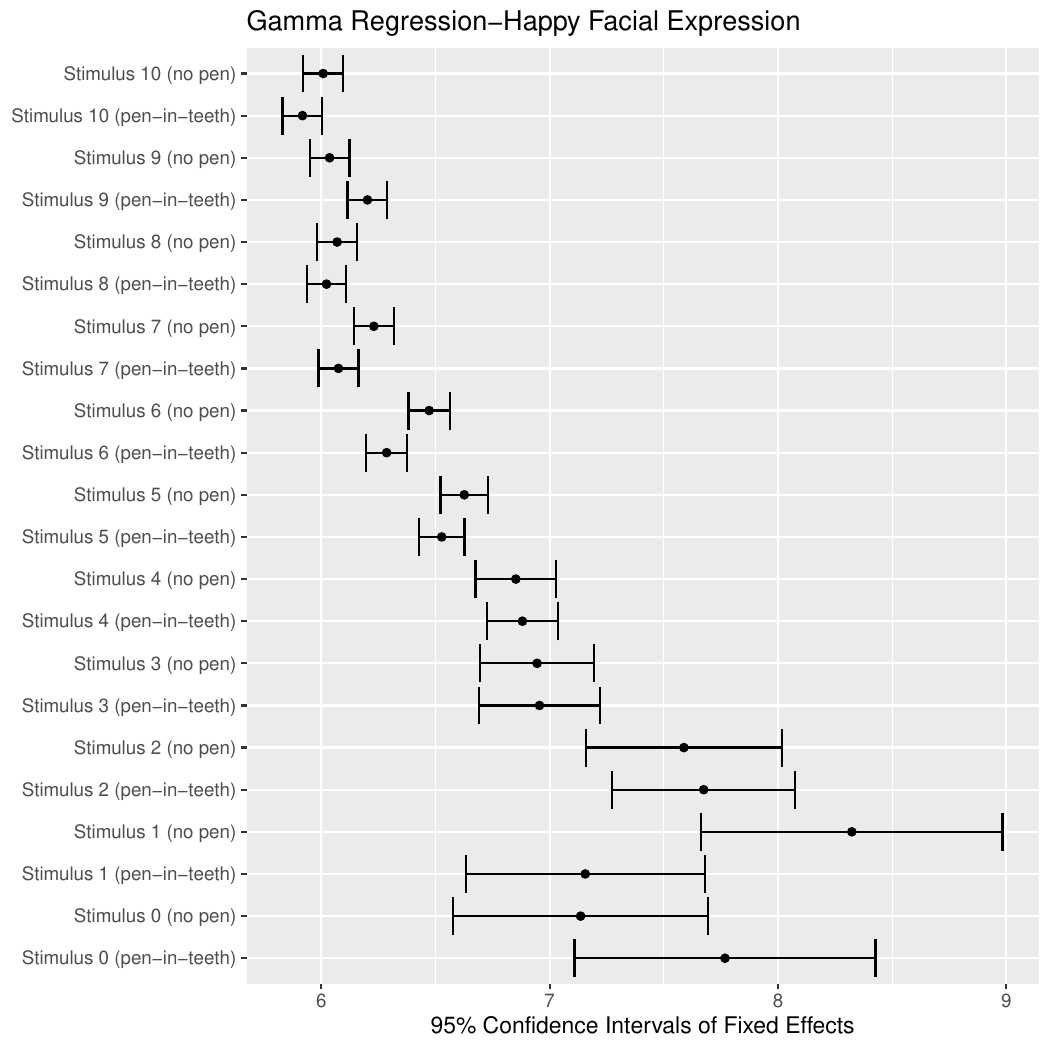}}
\caption{Fixed-effects estimation results under the Gamma model, displaying point estimates (dots) and their associated 95\% confidence intervals for all stimulus combinations. The figure is divided into two panels: the left panel presents results for ``sad'' facial expression responses, while the right panel shows results for ``happy'' facial expression responses. For ``sad'' responses, the estimated random-effect variance per individual ($\hat{\tau^2}$) was approximately 0.21, with an estimated dispersion $\hat{\phi}_G\approx 0.84$. For ``happy'' responses, the random-effect variance was approximately 0.15, with $\hat{\phi}_G\approx 0.83$ (see Eq. \protect\eqref{eq:disp-G} below for the corresponding expressions in the [estimated] marginal variance). Non-overlap of 95\% confidence intervals strongly suggests significance, but overlap does not rule it out (the correct approach is to compute the confidence intervals for the difference or perform a hypothesis test) [\protect\shortciteA{schenker2001judging}, \protect\shortciteA{wright2019primer}].}\label{fig:fixed-effects-gamma}
\end{figure}

The fixed-effects parameter vector $\beta$ in Equation \eqref{eq:hiestruc} consists of 22 components: $\beta=(\beta_1, ...,\beta_{22})^\top$. For the random effects, we considered 116 IID Normal variables $\{C_k\}_{k=1,...,116}$ with mean 0 and variance  $\tau^2$. Both IG and Gamma models employed the logarithmic link function. The parameters were estimated using Maximum Likelihood (ML) methods, implemented using the \textit{glmer} function from the \textit{lme4} package in \textbf{R}. This function computes ML estimates for GLMMs by approximating the log-likelihood through Gauss-Hermite quadrature and maximizing it via numerical optimization. The complete code and dataset are available at \url{https://cutt.ly/feXh490H}. The parameter estimates for the Gamma model are displayed in Fig. \ref{fig:fixed-effects-gamma}.

\begin{figure}[htbp]
\centering
\resizebox{6cm}{!}{\includegraphics{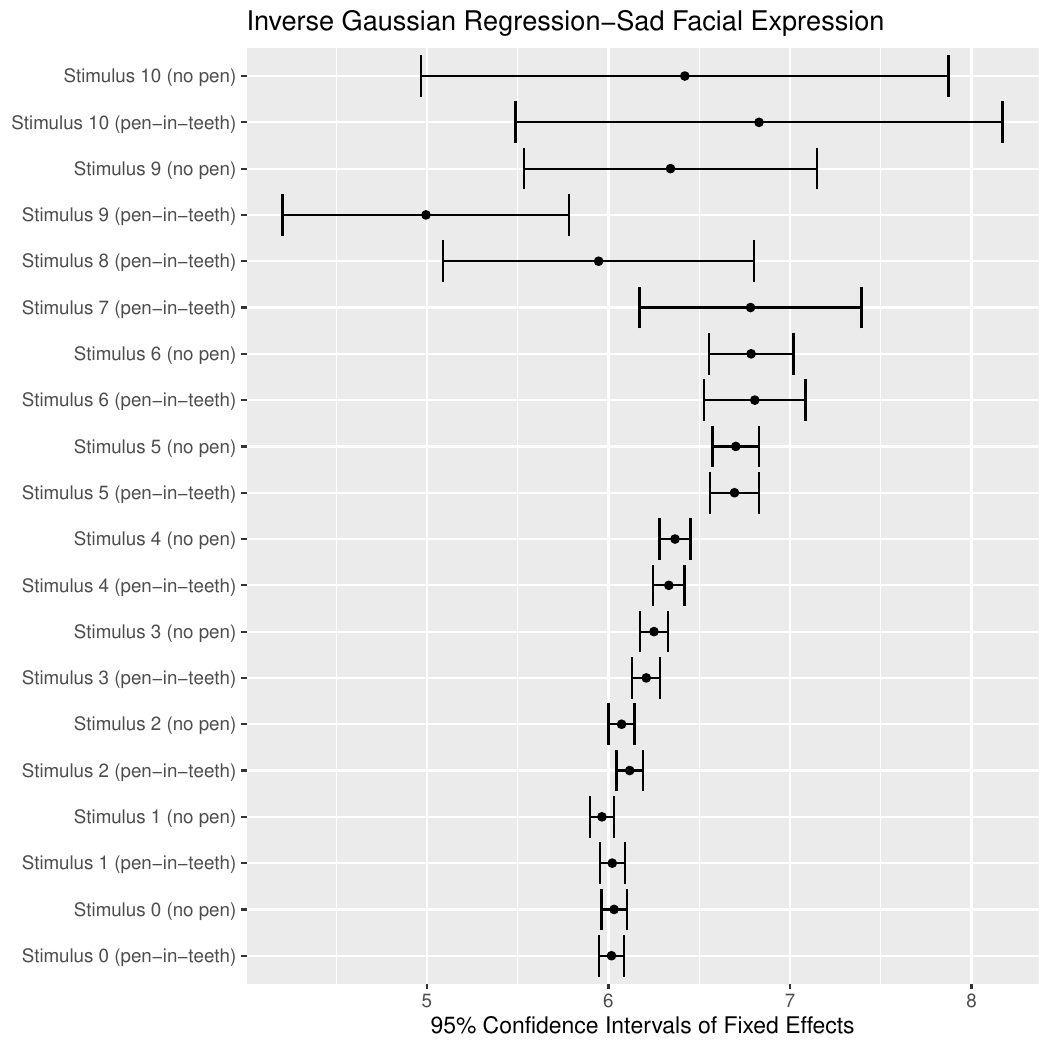}}
\resizebox{6cm}{!}{\includegraphics{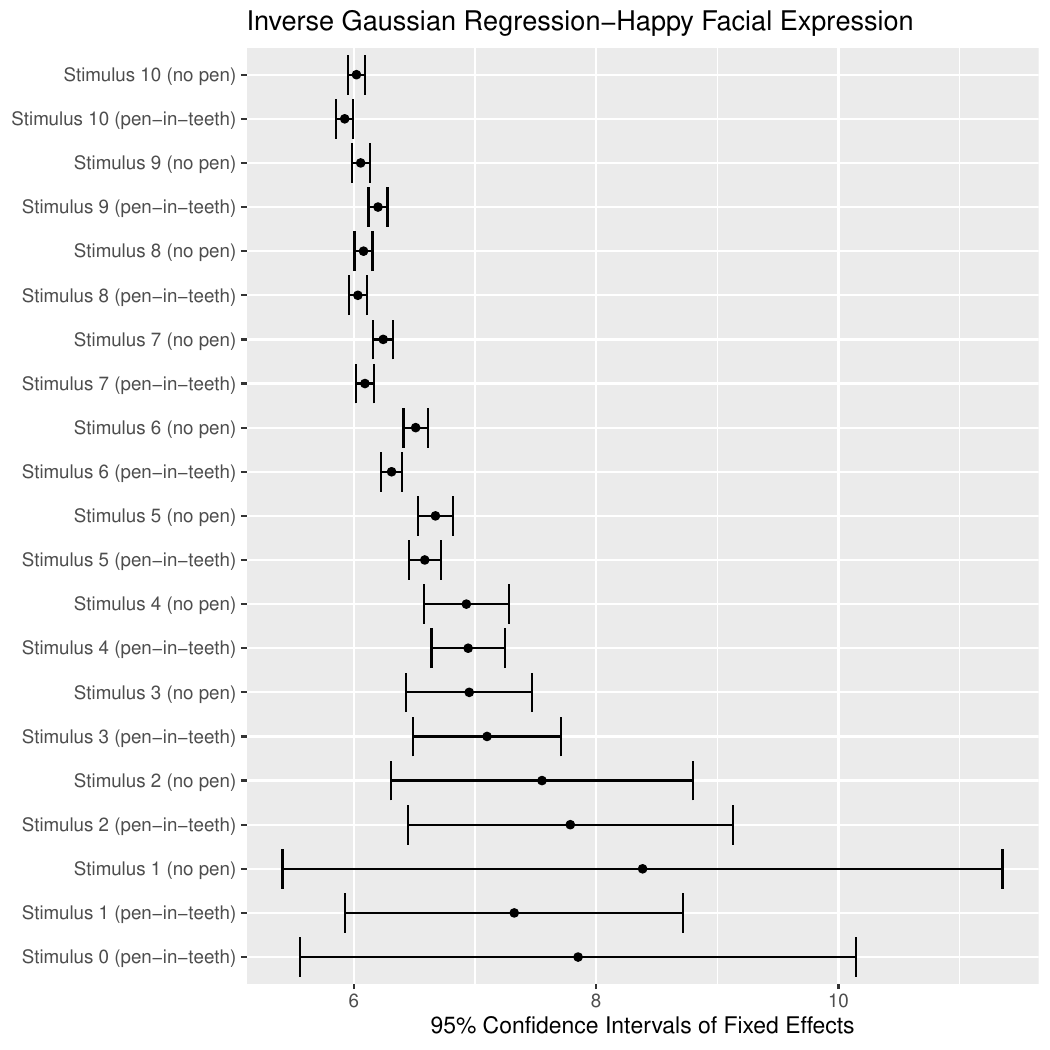}}
\caption{Fixed-effects estimation results under the IG model, displaying point estimates (dots) and their associated 95\% confidence intervals for most stimulus combinations. The figure is divided into two panels: the left panel presents results for ``sad'' facial expression responses, while the right panel shows results for ``happy'' facial expression responses. For ``sad'' responses, the estimated random-effect variance per individual ($\hat{\tau^2}$) was approximately 4.81, with an estimated dispersion $\hat{\phi}_{IG}\approx 0.0022$. For ``happy'' responses, the random-effect variance was approximately 1.65, with $\hat{\phi}_{IG}\approx 0.0026$ (see Eq. \protect\eqref{eq:disp-IG} below for the corresponding expressions in the [estimated] marginal variance). Note that confidence intervals could not be obtained for all stimulus combinations due to numerical convergence issues (specifically, for stimuli 7 and 8 under condition ``no pen'', given ``Sad Facial Expression'' response; and stimulus 0 under condition ``no pen'',  given ``Happy Facial Expression'' response). Non-overlap of 95\% confidence intervals strongly suggests significance, but overlap does not rule it out (the correct approach is to compute the confidence intervals for the difference or perform a hypothesis test) [\protect\shortciteA{schenker2001judging}, \protect\shortciteA{wright2019primer}].}\label{fig:fixed-effects-ig}
\end{figure}

Due to convergence issues in the iterative numerical method used to calculate estimation errors, we were unable to obtain all the confidence intervals for the IG model (this will be discussed in detail in Section \ref{sec:discussion}). Figure \ref{fig:fixed-effects-ig} therefore presents only those fixed-effects estimates for which we successfully computed the estimation errors.

Our analysis reveals a consistent pattern in RTs across both emotional classifications. As shown in Figs. \ref{fig:fixed-effects-gamma} and \ref{fig:fixed-effects-ig}, the left plots demonstrate that participants who correctly identified unambiguous ``sad'' faces (stimulus 0 and 1) exhibited faster RTs with minimal variance. In contrast, when participants misclassified clearly ``happy'' faces (stimulus 9 and 10) as ``sad'', their RTs were notably slower with increased variability, though this pattern was observed in only a small subset of participants. The right plots in Figs. \ref{fig:fixed-effects-gamma} and \ref{fig:fixed-effects-ig} present a symmetrical pattern: correct classification of unambiguous ``happy'' faces (stimulus 9 and 10) resulted in faster RTs with low variability, while misclassification of clearly ``sad'' faces (stimulus 0 and 1) as ``happy'' led to slower RTs with higher variability, again observed in a limited number of participants.

The Akaike Information Criterion (AIC) values were as follows: 126498.837 and 126122.440 for the Gamma model under ``sad'' and ``happy'' responses, respectively, and 149001.247 and 149739.235 for the IG model under ``sad'' and ``happy'' responses, respectively. These results indicate that, although the IG model has a smaller  dispersion ($\hat{\phi}_{IG}\approx 0.002$ versus $\hat{\phi}_{G}\approx 0.8$, under both ``sad'' and ``happy'' responses), it exhibits a much larger between-individual random-effect variance ($\approx 4.81$ and $\approx 1,65$ for IG, under  under ``sad'' and ``happy'' responses respectively, versus $\approx 0.21$ and $\approx 0.15$ for Gamma,  under ``sad'' and ``happy'' responses respectively); on the likelihood/AIC scale, the Gamma model provides the superior fit.

To address the choice of random-effect structure, Table~\ref{tab:modelcmp} compares the random-intercept models used above with their counterparts that add a participant-level random slope for the stimulus score. For the Gamma family the AIC favours the random-slope model (lower by $245$ for ``sad'' and $230$ for ``happy''), even though the additional slope variance is small ($\approx 0.01$); however, the fixed-effect estimates are very highly correlated between the two structures ($0.97$ and $0.99$) and differ only marginally on average (mean absolute difference $\le 0.07$ on the log scale), so the reconstructed diffusion-process parameters reported above are essentially unchanged. For the IG family the AIC instead favors the random-intercept model, and the random-slope fits are numerically degenerate (random-effect correlation at the $\pm 1$ boundary) with unstable fixed effects. We therefore retain the random-intercept specification as our main model---for interpretability and for comparability across families---while noting transparently that the Gamma random-slope model is marginally preferred by AIC without altering any substantive conclusion.

\begin{table}[htbp]
\centering
\small
\setlength{\tabcolsep}{4pt}
\caption{Comparison of random-effect structures: random intercept (by participant) versus random intercept $+$ random slope for the stimulus score, for the Gamma and IG GLMMs under each response. $k$ is the number of estimated parameters; $\sigma^2_{\text{int}}$, $\sigma^2_{\text{slope}}$ and $\rho$ are the random-intercept variance, the random-slope variance and their correlation, respectively. Lower AIC within each family/response pair in bold. All models converged and none was flagged singular by \texttt{lme4}; the IG random-slope models reach the $\rho=\pm1$ boundary, indicating a degenerate random-effect covariance.}
\label{tab:modelcmp}
\small
\begin{tabular}{lllccccccc}
\hline
Family & Resp. & RE structure & $k$ & $\log L$ & AIC & BIC & $\sigma^2_{\text{int}}$ & $\sigma^2_{\text{slope}}$ & $\rho$ \\
\hline
Gamma & sad   & intercept       & 24 & $-63225.4$ & 126498.8          & 126668.9 & 0.214 & ---   & ---     \\
Gamma & sad   & intercept+slope & 26 & $-63100.9$ & \textbf{126253.8} & 126438.0 & 0.215 & 0.011 & $-0.29$ \\
Gamma & happy & intercept       & 24 & $-63037.2$ & 126122.4          & 126292.5 & 0.153 & ---   & ---     \\
Gamma & happy & intercept+slope & 26 & $-62920.3$ & \textbf{125892.5} & 126076.8 & 0.913 & 0.010 & $-0.92$ \\
IG    & sad   & intercept       & 24 & $-74476.6$ & \textbf{149001.2} & 149171.3 & 4.809 & ---   & ---     \\
IG    & sad   & intercept+slope & 26 & $-74475.8$ & 149003.5          & 149187.8 & 4.195 & 0.005 & $1.00$  \\
IG    & happy & intercept       & 24 & $-74845.6$ & \textbf{149739.2} & 149909.3 & 1.651 & ---   & ---     \\
IG    & happy & intercept+slope & 26 & $-74845.5$ & 149743.0          & 149927.3 & 4.466 & 0.010 & $-1.00$ \\
\hline
\end{tabular}
\end{table}

Given that the link function is $\log$ and the $C_k$'s are IID as normal with a mean of 0 and variance $\tau^2$, we can derive the $\mu_i$'s using Equation \eqref{eq:mean} in our specific application:

\[
\mu_i=\int \exp\left(x_{ij}^\top\beta+c\right)\phi_{0,\tau^2}(c)dc=\int \exp\left(\beta_i+c\right)\phi_{0,\tau^2}(c)dc
\]
\[
=\exp\left(\beta_i+\frac{\tau^2}{2}\right),
\]
where $\phi_{0,\tau^2}(\cdot)$ is the probability density of a $N(0,\tau^2)$. Then, we can estimate the (marginal) expectation of the $Y_{ijk}$'s as:
\[
\hat{\mu}_i=\exp\left(\hat{\beta}_i+\frac{\hat{\tau^2}}{2}\right).
\]

For the marginal variance expressed in Equation \eqref{eq:var}, we can derive its components as follows: 
the term $\int E(Y_{ijk}|c)^2\phi_{0,\tau^2}(c)dc$ equals $\exp\left(2\beta_i+2\tau^2\right)$, while $\mu_i^2$ is equal to $\exp\left(2\beta_i+\tau^2\right)$. The integral $\int V(Y_{ijk}|c)\phi_{0,\tau^2}(c)dc$ will depend on the relationship \eqref{eq:mean-var} for each case: for the Gamma distribution, $V(Y_{ijk}|c)=\phi_G\mu_{ijk}^2=\phi_G\exp\{2\beta_i+2c\}$, and then  $\int V(Y_{ijk}|c)\phi_{0,\tau^2}(c)dc=\phi_G\exp\{2\beta_i+2\tau^2\}$, finally resulting $\sigma^2_i=\phi_G\exp\{2\beta_i+2\tau^2\}+\exp\left(2\beta_i+2\tau^2\right)-\exp\left(2\beta_i+\tau^2\right)$. And for the IG distribution, $V(Y_{ijk}|c)=\phi_{IG}\mu_{ijk}^3=\phi_{IG}\exp\{3\beta_i+3c\}$, and then  $\int V(Y_{ijk}|c)\phi_{0,\tau^2}(c)dc=\phi_{IG}\exp\{3\beta_i+(9/2)\tau^2\}$, finally resulting $\sigma^2_i=\phi_{IG}\exp\{3\beta_i+(9/2)\tau^2\}+\exp\left(2\beta_i+2\tau^2\right)-\exp\left(2\beta_i+\tau^2\right)$.

So, the estimator of $\sigma_i^2$ depends on the family. For the Gamma model,
\begin{equation}\label{eq:disp-G}
\hat{\sigma}_i^2=\hat{\phi}_G\exp\!\left(2\hat{\beta}_i+2\hat{\tau^2}\right)+\exp\!\left(2\hat{\beta}_i+2\hat{\tau^2}\right)-\exp\!\left(2\hat{\beta}_i+\hat{\tau^2}\right),
\end{equation}
and for the IG model,
\begin{equation}\label{eq:disp-IG}
\hat{\sigma}_i^2=\hat{\phi}_{IG}\exp\!\left(3\hat{\beta}_i+\tfrac{9}{2}\hat{\tau^2}\right)+\exp\!\left(2\hat{\beta}_i+2\hat{\tau^2}\right)-\exp\!\left(2\hat{\beta}_i+\hat{\tau^2}\right).
\end{equation}

With these components established, we can now reconstruct the underlying cognitive processes for each stimulus combination and response type, following the methodology outlined in Section \ref{sec:par-map}. 

\textit{Remark.} On the one hand, the explicit derivation of expressions \eqref{eq:mean}-\eqref{eq:var} was made possible by our specific model choices: the logarithmic link function $h(\cdot)=\log(\cdot)$ and normally distributed random-effects. For alternative specifications of these components, these expressions may not have analytical solutions and would require numerical approximation methods. And, on the other hand, and as previously mentioned, when  the distribution of the RTs (conditioned on each response) induced by our Gamma/IG GLMM [i.e., from Equation \eqref{eq:hiestruc}], is integrated over the random effects (i.e., marginalizing them), their resulting distribution  are not exactly a Gamma or exactly an IG distributions, but rather a mixture, whose mixing distribution is given through the random-effect normal density $\phi_{0,\tau^2}(\cdot)$. So, our reconstruction for the corresponding diffusion processes should be interpreted as a `moment-based approximation, where such a reconstruction is exact at $C=0$ (see next Subsection).

\subsection{Reconstructed
diffusion-process parameters in the empirical example}

Following Section~\ref{sec:par-map}, we reconstruct the diffusion-process parameters at the conditional level $c=0$ (a typical individual), where $Y_{ijk}\mid c=0$ is exactly IG (resp.\ Gamma) and the mapping is therefore exact. Because the dispersion is constant, the IG starting point $\hat{a}_i=1/\sqrt{\hat{\phi}_{IG}}$ and the Gamma shape $\hat{\alpha}_i=1/\hat{\phi}_{G}$ do not depend on the stimulus level: the stimulus-driven variation is carried entirely by the drift rate $\hat{\nu}_i$. Tables~\ref{tab:recon-ig} and~\ref{tab:recon-gamma} report $\hat{\nu}_i$ together with $95\%$ delta-method intervals for representative levels, separately for the two responses and the two conditions. Since the Gamma model attains a substantially lower AIC than the IG model (Section~\ref{sec:experiment}), we regard the Gamma reconstruction as the preferred one and report the IG results for completeness.

\begin{table}[htbp]
\centering
\caption{Reconstructed IG diffusion-process parameters at the conditional level $c=0$ (typical individual), for representative stimulus levels. The starting point $\hat{a}_i=1/\sqrt{\hat{\phi}_{IG}}$ is constant within each response ($\approx 21.3$ for ``sad'', $\approx 19.8$ for ``happy''); only the drift rate $\hat{\nu}_i$ varies across stimuli. Brackets give 95\% delta-method intervals. The cell ``happy/no-pen/stimulus~0'' is degenerate (too few ``happy'' responses to a fully-sad face) and is omitted.}
\label{tab:recon-ig}
\begin{tabular}{llcrcc}
\hline
Response & Condition & Stimulus & $\hat{\mu}_{c=0}$ (ms) & $\hat{\nu}$ & 95\% CI \\
\hline
sad & pen-in-teeth & 0 & 410 & 0.052 & [0.048,\,0.056] \\
 & pen-in-teeth & 1 & 412 & 0.052 & [0.048,\,0.055] \\
 & pen-in-teeth & 4 & 563 & 0.038 & [0.035,\,0.041] \\
 & pen-in-teeth & 5 & 808 & 0.026 & [0.023,\,0.030] \\
 & pen-in-teeth & 6 & 903 & 0.024 & [0.018,\,0.031] \\
 & pen-in-teeth & 9 & 148 & 0.144 & [0.066,\,0.318] \\
 & pen-in-teeth & 10 & 925 & 0.023 & [0.006,\,0.088] \\
\hline
 & no-pen & 0 & 416 & 0.051 & [0.048,\,0.055] \\
 & no-pen & 1 & 389 & 0.055 & [0.051,\,0.059] \\
 & no-pen & 4 & 582 & 0.037 & [0.034,\,0.040] \\
 & no-pen & 5 & 814 & 0.026 & [0.023,\,0.030] \\
 & no-pen & 6 & 885 & 0.024 & [0.019,\,0.030] \\
 & no-pen & 9 & 568 & 0.038 & [0.017,\,0.084] \\
 & no-pen & 10 & 614 & 0.035 & [0.008,\,0.148] \\
\hline
happy & pen-in-teeth & 0 & 2568 & 0.008 & [0.001,\,0.076] \\
 & pen-in-teeth & 1 & 1516 & 0.013 & [0.003,\,0.053] \\
 & pen-in-teeth & 4 & 1036 & 0.019 & [0.014,\,0.026] \\
 & pen-in-teeth & 5 & 724 & 0.027 & [0.024,\,0.031] \\
 & pen-in-teeth & 6 & 551 & 0.036 & [0.033,\,0.039] \\
 & pen-in-teeth & 9 & 492 & 0.040 & [0.037,\,0.043] \\
 & pen-in-teeth & 10 & 374 & 0.053 & [0.049,\,0.057] \\
\hline
 & no-pen & 1 & 4374 & 0.005 & [0.000,\,0.088] \\
 & no-pen & 4 & 1021 & 0.019 & [0.014,\,0.028] \\
 & no-pen & 5 & 791 & 0.025 & [0.022,\,0.029] \\
 & no-pen & 6 & 672 & 0.029 & [0.027,\,0.033] \\
 & no-pen & 9 & 427 & 0.046 & [0.043,\,0.050] \\
 & no-pen & 10 & 412 & 0.048 & [0.045,\,0.052] \\
\hline
\end{tabular}
\end{table}

\begin{table}[htbp]
\centering
\caption{Reconstructed Gamma diffusion-process parameters at the conditional level $c=0$, for representative stimulus levels. The shape $\hat{\alpha}_i=1/\hat{\phi}_{G}$ is constant within each response ($\approx 1.19$ for ``sad'', $\approx 1.21$ for ``happy''); the drift $\hat{\nu}_i=(\hat{\beta}^{\rm sc}_i/2)^{-1/2}$ with $\hat{\beta}^{\rm sc}_i=\hat{\phi}_{G}\exp(\hat{\beta}_i)$ varies across stimuli. Brackets give 95\% delta-method intervals.}
\label{tab:recon-gamma}
\begin{tabular}{llcrcc}
\hline
Response & Condition & Stimulus & $\hat{\mu}_{c=0}$ (ms) & $\hat{\nu}$ & 95\% CI \\
\hline
sad & pen-in-teeth & 0 & 424 & 0.075 & [0.071,\,0.079] \\
 & pen-in-teeth & 1 & 424 & 0.075 & [0.071,\,0.079] \\
 & pen-in-teeth & 4 & 579 & 0.064 & [0.061,\,0.067] \\
 & pen-in-teeth & 5 & 841 & 0.053 & [0.050,\,0.056] \\
 & pen-in-teeth & 6 & 952 & 0.050 & [0.045,\,0.055] \\
 & pen-in-teeth & 9 & 166 & 0.120 & [0.063,\,0.228] \\
 & pen-in-teeth & 10 & 1076 & 0.047 & [0.030,\,0.074] \\
\hline
 & no-pen & 0 & 424 & 0.075 & [0.071,\,0.079] \\
 & no-pen & 1 & 401 & 0.077 & [0.073,\,0.081] \\
 & no-pen & 4 & 598 & 0.063 & [0.060,\,0.066] \\
 & no-pen & 5 & 837 & 0.053 & [0.050,\,0.056] \\
 & no-pen & 6 & 935 & 0.050 & [0.046,\,0.055] \\
 & no-pen & 9 & 568 & 0.065 & [0.047,\,0.089] \\
 & no-pen & 10 & 521 & 0.067 & [0.040,\,0.114] \\
\hline
happy & pen-in-teeth & 0 & 2364 & 0.032 & [0.021,\,0.048] \\
 & pen-in-teeth & 1 & 1282 & 0.043 & [0.032,\,0.060] \\
 & pen-in-teeth & 4 & 973 & 0.050 & [0.045,\,0.055] \\
 & pen-in-teeth & 5 & 684 & 0.059 & [0.056,\,0.063] \\
 & pen-in-teeth & 6 & 537 & 0.067 & [0.064,\,0.070] \\
 & pen-in-teeth & 9 & 494 & 0.070 & [0.067,\,0.073] \\
 & pen-in-teeth & 10 & 372 & 0.081 & [0.077,\,0.084] \\
\hline
 & no-pen & 0 & 1256 & 0.044 & [0.031,\,0.062] \\
 & no-pen & 1 & 4118 & 0.024 & [0.016,\,0.036] \\
 & no-pen & 4 & 946 & 0.051 & [0.046,\,0.056] \\
 & no-pen & 5 & 755 & 0.057 & [0.053,\,0.060] \\
 & no-pen & 6 & 647 & 0.061 & [0.058,\,0.064] \\
 & no-pen & 9 & 418 & 0.076 & [0.073,\,0.080] \\
 & no-pen & 10 & 407 & 0.077 & [0.074,\,0.081] \\
\hline
\end{tabular}
\end{table}

Two patterns emerge consistently across both families. First, the drift rate increases with the emotional clarity of the stimulus in the direction of the response: for ``sad'' responses $\hat{\nu}_i$ is largest for the clearly-sad stimuli ($0$--$1$) and decreases towards the happy end, whereas for ``happy'' responses the pattern is reversed, with the largest drift for the clearly-happy stimuli ($9$--$10$). This is consistent with faster, more decisive evidence accumulation when the facial expression is unambiguous. Second, drift estimates for responses that contradict the stimulus (e.g.\ a ``sad'' response to a clearly-happy face) rest on few trials and carry correspondingly wide intervals. The pen-in-teeth and no-pen conditions yield broadly similar drift profiles, and within-stimulus differences are small relative to the intervals; the data thus provide only weak evidence that holding a pen modifies the drift, and none that it shifts the (level-independent) starting point or boundary. The reported intervals propagate the sampling variability of the fixed effects $\hat{\beta}_i$ through the delta method; the dispersion $\hat{\phi}$---and hence the constant $\hat{a}_i$ and $\hat{\alpha}_i$---can be additionally assessed by a parametric bootstrap.

\begin{figure}[htbp]
\centering
\includegraphics[width=\textwidth]{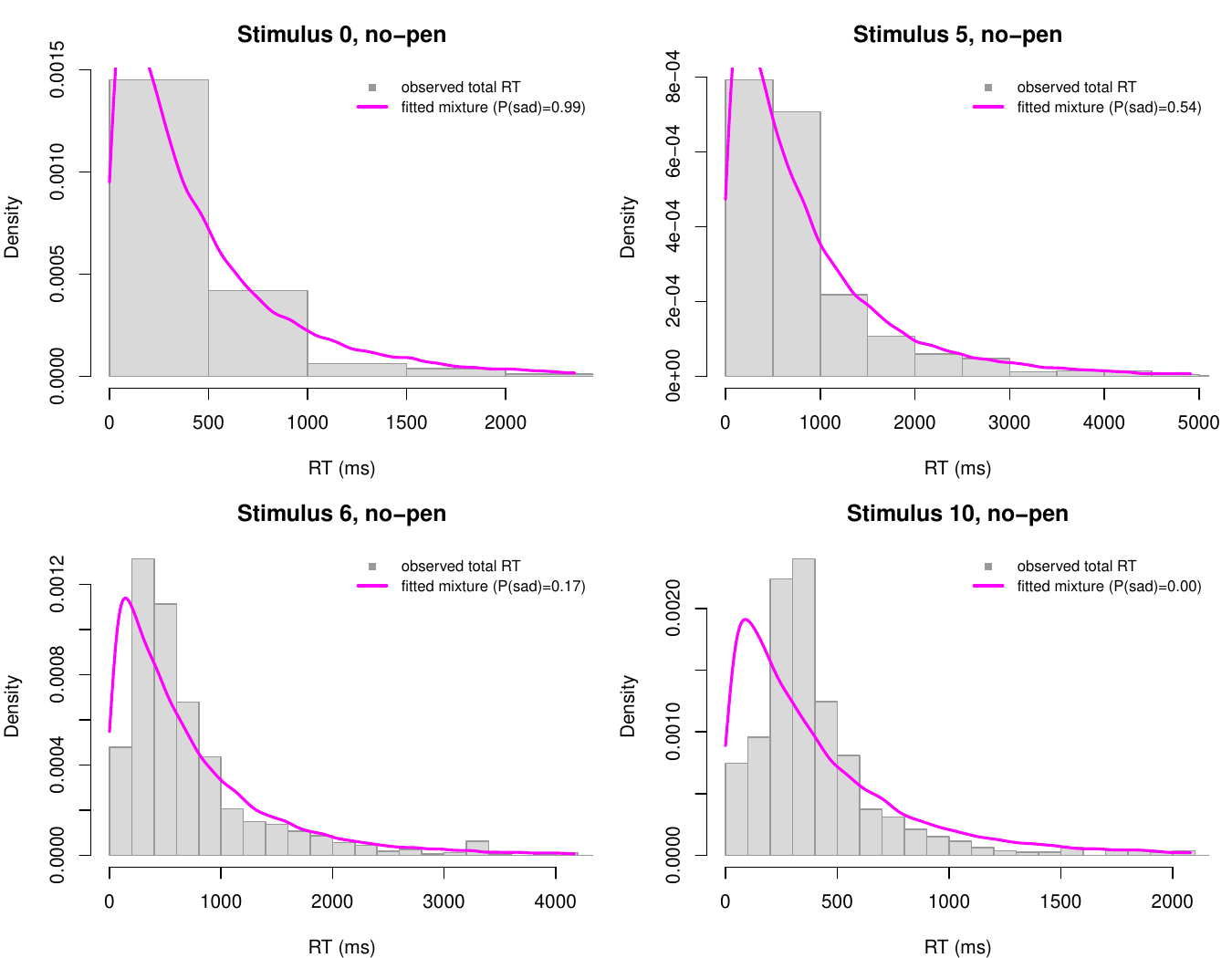}
\caption{Observed total RT distribution (grey histogram) and fitted response mixture $\hat{P}(R_1\mid i)\hat{f}_i(y\mid R_1)+\hat{P}(R_2\mid i)\hat{f}_i(y\mid R_2)$ (magenta), under the Gamma model, for four stimulus levels (no-pen condition) spanning the sad--happy continuum: clearly sad (0), ambiguous (5, $\hat{P}(\text{sad})\approx0.54$), mostly happy (6), and clearly happy (10). Per-response marginal densities are obtained by simulating the fitted GLMM; weights are empirical response frequencies.}
\label{fig:mixture}
\end{figure}

Finally, to illustrate the law of total probability that links the response-conditional RT distributions to the marginal one, Figure~\ref{fig:mixture} overlays, for representative stimulus levels along the sad--happy continuum, the fitted response mixture
\begin{equation}\label{eq:frequencies}
\hat{f}_i(y)=\hat{P}(R_1\mid i)\,\hat{f}_i(y\mid R_1)+\hat{P}(R_2\mid i)\,\hat{f}_i(y\mid R_2)
\end{equation}
on the observed total RT distribution, using the Gamma model. The per-response densities $\hat{f}_i(\cdot\mid R_l)$ are the marginal (random-effect-integrated) densities, obtained by simulating the fitted GLMM, and the weights $\hat{P}(R_l\mid i)$ are the empirical response frequencies. The mixture reproduces the observed distribution well across the continuum---including the ambiguous stimulus~5, where both responses carry substantial weight and RTs are markedly slower. The mild excess mass that the fitted densities place at very short RTs reflects the absence of a non-decision component ($\theta=0$); incorporating $\theta>0$ (Section~\ref{sec:discussion}) would shift the densities rightward.

\section{Discussion and scopes}\label{sec:discussion}

This paper demonstrates a relatively straightforward method for connecting two methodological approaches to modeling  RTs. The first approach models RTs as the result of the distribution of the first-hitting time of a simple diffusion process, simulating the cognitive processing underlying the choice of a response in a  choice task. The second approach utilizes a (conditional) GLMM framework. The advantage of this connection is twofold. First, starting with the diffusion model approach, theoretical models can be used to inform the selection of specific GLMM families. Second, after estimating the parameters within these GLMM families, conditioned on each response, we can reconstruct a simplified model of the underlying cognitive process that generated those responses, under a \textit{moment-based approximation}. 

While the presented approach involves certain simplifications, it offers potential for extension and generalization. For instance, within the GLMM framework, we could incorporate additional random effects to account for correlations within individual responses, and explore alternative link functions and transformations, as discussed in \shortciteA{lo2015transform}. Furthermore, the initial diffusion model describing the cognitive process behind a specific response included a ``non-decision'' parameter, which was ultimately set to zero here. Investigating the estimation of this parameter through a revised GLMM formulation—by considering shifted distributions with a hierarchical structure on $\theta$—would be a valuable direction for future research.

It is important to note that our focus was on modeling RTs conditional on the responses, rather than on the responses themselves. While a simplified approach to modeling response probabilities is presented in Equation \eqref{eq:responsesprobabs} and empirically calculated in \eqref{eq:frequencies} under our considered dataset, a more rigorous treatment is warranted. This is because responses are also influenced by fixed effects at different stimulus levels and by random effects associated with individuals, as highlighted in previous research [\shortciteA{van2007hierarchical}; \shortciteA{moscatelli2012modeling}; \shortciteA{molenaar2015bivariate}; \shortciteA{ranger2020modeling}]. Hence, modeling the responses themselves deserves its own attention, which may be addressed in future work.

Estimating GLMMs via ML typically requires numerical quadrature methods, as a closed-form analytical solution is unavailable. Popular choices include the Laplace approximation and adaptive Gauss-Hermite quadrature. However, these methods have been shown to produce biased estimators in certain contexts [\shortciteA{GH-biblio1}], and alternative approaches may be superior for GLMMs [\shortciteA{GH-biblio2}]. Stochastic Expectation-Maximization (EM) algorithms, such as the Stochastic Approximation EM (SAEM) algorithm [\shortciteA{delyon:1999}], offer an alternative to log-likelihood approximation methods. The SAEM algorithm has shown promising performance with complex mixed models like GLMMs, while retaining the advantages of the EM framework [\shortciteA{savic2009performance}; \shortciteA{SAEM:probit}]. Therefore, a worthwhile future direction would be to explore the application of the SAEM algorithm for estimating the model presented in this work.

We believe that our approach can be extended to other variants of GLMs, particularly the Double Hierarchical GLM (DHGLM). Despite its significant potential, the DHGLM, a sophisticated extension of GLMs, has not yet been applied to RT experiments. Its primary innovation is the inclusion of random effects in both the mean and dispersion components of the model. This dual inclusion allows for the simultaneous modeling of heteroscedasticity between clusters and heterogeneity in means, offering a more comprehensive analytical framework. A key advantage of the DHGLM is its use of $h$-likelihood, which provides a unified framework for model fitting through a single algorithm, thus avoiding the need for complex numerical integration or specifications of prior probabilities, leading to computational efficiency and practical implementation [\shortciteA{lee2006double}]. We propose that our conditional GLMM approach could be integrated into the DHGLM framework, leveraging the mean and shape parameters of the IG distribution and the shape and scale parameters of the Gamma distribution. This integration would facilitate more advanced analyses of RT responses, potentially yielding deeper insights into the underlying cognitive processes.

Considering the points discussed above and the demonstrated applicability shown in previous sections, the presented approach provides a solid foundation for further extensions. These extensions hold the potential to enhance the analysis and interpretation of experimental results, and broaden the applicability of the approach to multiple-choice task tests.

\subsubsection*{Data availability statement}

All data are available on link \url{https://cutt.ly/feXh490H}.

\subsubsection*{Acknowledgements}

This research was funded by the INICI-UV Grant (UVA 20993, 2021) and received additional support from Exploration-ANID (13220168, 2022). The authors extend their gratitude to Professor Javier Contreras, Professor Joaquín Cavieres, and Professor Raydonal Ospina for their invaluable insights and assistance with numerical implementations. We also thank anonymous reviewers, whose comments and advice have substantially improved our first version.
The published version of this mansucript can be found at \url{https://doi.org/10.1080/02664763.2026.2726564}

\subsubsection*{Disclosure Statement}

The authors declare that they have no conflicts of interest of any kind.

\bibliographystyle{apacitex}
\bibliography{references}

\end{document}